\documentclass[a4paper,11pt,headings=big,DIV=12]{article}
\pdfoutput=1
\usepackage{amsmath,amssymb,mathtools}
 
\usepackage[usenames, dvipsnames]{color}
\usepackage{tcolorbox} 
\definecolor{mygray}{gray}{0.2}
\usepackage{bbold}

\usepackage{jheppub}

\usepackage{bm,amssymb,slashed,graphicx,multirow,soul,mathtools,xspace,array}  
\usepackage{float}                   
\allowdisplaybreaks
\usepackage{ bbold }
\usepackage{subfigure}
\usepackage{slashed}
\usepackage{acronym}
   \usepackage{soul}

\definecolor{violet}{rgb}{0.94, 0.2, 0.8}
\definecolor{lightblue}{rgb}{0.39, 0.58, 0.93} 
\definecolor{asparagus}{rgb}{0.53, 0.66, 0.42}

\newcommand{\comA}[1]{ {#1}}

\newcommand{\EQ}{Eq.~}

\acrodef{PDG}[PDG]{Particle Data Group}
\acrodef{OPE}[OPE]{Operator Product Expansion}
\acrodef{FCNC}[FCNC]{flavour-changing neutral current}
\acrodef{RHC}[RHC]{right-handed currents}
\acrodef{SM}[SM]{Standard Model}
\acrodef{NP}[NP]{New Physics}
\acrodef{MFV}[MFV]{Minimal Flavour Violation}
\acrodef{SD}[SD]{short-distance}
\acrodef{LD}[LD]{long-distance}
\acrodef{DA}[DA]{distribution amplitude}

\newcommand{\vev}[1]{\langle #1 \rangle} 
\newcommand{\state}[1]{|#1\rangle}

\newcommand{\matel}[3]{\langle #1|#2|#3\rangle}
\newcommand{\al}{\alpha}
\newcommand{\be}{\beta}
\newcommand{\ga}{\gamma}

\newcommand{\de}{\delta}
\newcommand{\De}{\Delta}

\newcommand{\La}{\Lambda}

\newcommand{\sig}{ \sigma}

\newcommand{\TeV}{\,\mbox{TeV}}
\newcommand{\GeV}{\,\mbox{GeV}}
\newcommand{\MeV}{\,\mbox{MeV}}

\newcommand{\pl}{\!+\!}

\newcommand{\TAB}{Tab.~}
\newcommand{\FIG}{Fig.~}
\newcommand{\SEC}{Sec.~}

\newcommand{\APP}{App.~}
\newcommand{\APPs}{Apps.~}

\newcommand{\ORD}{{\cal O}}

\newcommand{\Gone}{G_1}
\newcommand{\Gtwo}{G_2}

\newcommand{\dil}{D}
\newcommand{\bary}{f}
\newcommand{\Tbary}{T^{(\bary)}}
\newcommand{\Tbaryud}[2]{T^{(\bary) #1 }_{#2}}
\newcommand{\scal}{\phi}
\newcommand{\ym}[1]{\frac{#1}{1+\ga_*}}
\newcommand{\mq}{m_q}
\newcommand{\dbar}{{d-1}}
\newcommand{\Tud}[2]{T^{#1}_{\phantom{#1} #2}}
\newcommand{\veci}[1]{{\mathbf{#1}}}
\newcommand{\Tphi}{T^{(\scal)}}
\newcommand{\Tphiud}[2]{T^{(\scal) #1 }_{#2}}
\newcommand{\mink}{\eta}

\definecolor{violet}{rgb}{0.94, 0.2, 0.8}
\definecolor{lightblue}{rgb}{0.39, 0.58, 0.93} 
\definecolor{lightgreen}{rgb}{0.1, 0.73, 0.33}

\DeclareOldFontCommand{\tt}{\normalfont\ttfamily}{\mathtt}

\newcommand*{\mathcolor}{}
\def\mathcolor#1#{\mathcoloraux{#1}}
\newcommand*{\mathcoloraux}[3]{%
  \protect\leavevmode
  \begingroup
    \color#1{#2}#3%
  \endgroup
}

\begin{document}

\title{\boldmath Dilaton and Massive Hadrons in a Conformal Phase}

\author[1]{Luigi Del Debbio,}
\author[1]{Roman Zwicky,}

\affiliation[1]{Higgs Centre for Theoretical Physics, School of Physics and
Astronomy, The University of Edinburgh, 
Peter Guthrie Tait Road, Edinburgh EH9 3FD, Scotland, UK}
\emailAdd{roman.zwicky@ed.ac.uk}
\emailAdd{luigi.del.debbio@ed.ac.uk}

\abstract{As the number of fermion fields is increased, gauge theories are
expected to undergo a transition from a QCD-like phase, characterised by
confinement and chiral symmetry breaking, to a conformal phase, where the theory
becomes scale-invariant at large distances. In this paper, we discuss some
properties of a third phase, where spontaneously broken conformal symmetry is
characterised by its Goldstone boson, the dilaton. In this phase, which we refer to as conformal dilaton phase,  
the massless pole corresponding to the Goldstone boson guarantees that the
conformal Ward identities are satisfied in the infrared despite the other
hadrons carrying mass. In particular, using renormalisation group arguments in
Euclidean space, we show that for massless quarks the trace of the energy
momentum tensor vanishes on all physical states as a result of the fixed point.
This implies the vanishing of the gluon condensate and suggests that the scale
breaking is driven by the quark condensate which has implications for the
cosmological constant. In addition form factors obey an exact constraint for
every hadron and are thus suitable probes to identify this phase in the context
of lattice Monte Carlo studies. For this purpose we examine how the system
behaves under explicit symmetry breaking, via quark-mass and finite-volume
deformations. The dilaton mass shows hyperscaling under mass deformation,
viz. $m_{D} = {\cal O}(m_q^{1/(1+\gamma^*)})$. This provides another clean
search pattern.} 
\maketitle

\flushbottom


\setcounter{tocdepth}{3}
\setcounter{page}{1}
\pagestyle{plain}

\section{Introduction}
\label{sec:Intro}

It is well-known, since the seminal work of Ref.~\cite{Banks:1981nn}, that gauge
theories in $d=4$ show very different infrared (IR) behaviour depending on the
matter representation, the number of  flavours $N_f$ and colours $N_c$. As the
matter content is varied, these theories undergo a transition between a QCD-like
phase where chiral symmetry is spontaneously broken, and hadron confinement
takes place, and a phase where conformal symmetry is exhibited by the scaling of
the correlation functions in the IR.\footnote{It is generally believed that
confinement and chiral symmetry go hand in hand because the pion match the
anomaly of the quark in the IR. In some supersymmetric gauge theories the role
of the pions can be taken by massless baryons and this phase is referred to as
s-confinement \cite{Seiberg:1994bz,Terning:2006bq}.} The latter phase is
referred to  as the ``conformal window". Recent results are summarised in
Ref.~\cite{Nogradi:2016qek}.

In this work we would like to investigate some properties of a third phase where
conformal symmetry is spontaneously broken, leading to the appearance of a
Goldstone boson (GB), the dilaton.\footnote{Throughout we will not distinguish
conformal and scale (dilatation) invariance. It is widely believed that scale
invariance implies conformal invariance in a wide class of theories in four
dimensions, see e.g. Ref.~\cite{Nakayama:2013is} for a review.}  
The dilaton has been widely studied in the literature as a candidate model for a
composite version of the Higgs~\cite{Yamawaki:1985zg,Cata:2018wzl} with various
effective Lagrangians~\cite
{Appelquist:2020bqj,Appelquist:2017wcg,Appelquist:2017vyy,Golterman:2016lsd,Matsuzaki:2013eva},
or as a driving field theory version force of inflation~\cite{Csaki:2014bua}.
In this work we focus on the dilaton as the catalyst
to the massive hadronic spectrum; indeed the massless pole corresponding to the
dilaton allows for the conformal Ward identity (WI) to be satisfied even in the
presence of massive states in the spectrum. In particular the trace of the
energy momentum tensor (EMT) vanishes on physical states $\phi_i$,
$\matel{\phi_2}{\Tud{\mu}{\mu}(x)} {\phi_1} \to 0$ as shown in \SEC\ref{sec:WI}.

\subsection{Axial and Dilatation Ward Identities}
\label{sec:ADWI}

It is well-known that the pion decay constant $F_\pi$ is the order parameter of
spontaneous chiral symmetry breaking. The dilaton decay constant $F_{\dil}$
plays the analogous role for the spontaneous breaking of dilatation or scale
symmetry. It seems beneficial to treat them in parallel here. The decay
constants are defined as~\footnote{\label{foot:fD} The second equation below is
consistent with $\matel{0}{T_{\mu\nu}}{\dil(q)} =  
\frac{F_{\dil} m_{\dil}^2}{\dbar}  
(\mink_{\mu\nu} - q_\mu q_\nu/m_{\dil}^2)$ cf. also Eq.~\eqref{eq:dilWI}.} 
\begin{alignat}{3}
\label{eq:fdecay}
& \Gamma^{(ab)}_{5\mu}(q)  &\;=\;& 
  \matel{0}{J_{5\mu}^a(0)}{\pi^b(q)} &\;=\;&  
  i F_\pi  q_\mu  \de^{ab} \;, \quad  \nonumber \\[0.1cm]
& \Gamma_{\mu}(q)  &\;=\;& \matel{0}{J^{\dil}_\mu(0)}{\dil(q)} &\;=\;& 
  i {F_{\dil}} q_\mu \;,
\end{alignat} 
where the Noether currents associated to the broken symmetries are respectively
$J_{5\mu}^a(x) = \bar{q}(x) T^a \ga_\mu \ga_5 q(x)$ and $J^{\dil}_\mu(x) = x^\nu
T_{\mu\nu}(x)$, where $F_\pi \approx 92 \MeV$ in QCD and $T^a$ is a generator of
the broken axial flavour symmetry $SU(N_F)$. The divergences of the currents are
given by the explicit and anomalous symmetry breaking; using~\footnote{All our
conventions are  specified in \APP\ref{app:conv}. The trace anomaly 
\cite{Minkowski:1976en,Collins:1976yq,Nielsen:1977sy} 
contains
further equation of motion terms  which vanish on
physical states and are not of interest to our work.}
\begin{align}
  \partial \cdot J^a_{5}(x) &= 2 \mq  P^a(x) = 2 \mq \, \bar q(x) T^ai \ga_5  q (x) \nonumber  \; , \\
  \label{eq:DivNoetherCurrTwo}
  \partial \cdot J^D(x) &= \Tud{\mu}{\mu}(x) = 
    \frac{\be}{2 g} G^2(x) \pl \mq(1\pl \ga) \bar{q}(x) q(x) \; ,
\end{align}
one obtains
\begin{alignat}{4}
  \label{eq:chiWI} 
  & i^{-1} q^\mu \Gamma^{(ab)}_{5\mu}(0)  &\;=\;&
    2 \mq \matel{0}{P^a(0)} {\pi^b(q)} &\;=\;& 
       F_\pi  m_{\pi}^2  \de^{ab} &\;\stackrel{\textrm{sym}}{\to }\;& 0  \;, 
    \quad  \\[0.1cm]
  \label{eq:dilWI} 
    & i^{-1} q^\mu \Gamma_\mu(0)  &\;=\;&  
       \matel{0}{ \frac{\be}{2 g} G^2(0) \pl \mq(1\pl \ga) 
    \bar{q}q(0) }{\dil(q)}  &\;=\;&
    {F_{\dil}} m_{\dil}^2  &\;\stackrel{\textrm{sym}}{\to }\;&    0 \;.
\end{alignat}
These equations vanish in the symmetry limit $\mq \to 0$. 
For the dilaton WI \eqref{eq:dilWI} this is not obvious as there is anomalous breaking of scale symmetry in addition. However 
in \SEC\ref{sec:WI} we prove, using renormalisation group (RG) arguments in Euclidean space, that the equation holds.
\begin{table}[h]
	\centering
	\begin{tabular}{ l   |  l l   ||  l   l   }
   \text{particle} 	  &    \text{non-Goldstone}    &  ($\frac{\eta_{F_{\dil',\pi'}} }{1+\ga_*} \geq 1$)    &   \text{Goldstone}    &
    \\  \hline
    $J^{PC} = 0^{-+}$     &  $m_{\pi'} = \ORD(\La)$    &    $\comA{F_{\pi'}} =  \ORD(\mq^{\frac{\eta_{F_{\pi'}} }{1+\ga_*}} )$
  &  $m_{\pi} =  \ORD(\mq^\ym{1})$    &    $F_{\pi} = \ORD(\La)$
    \\ 
    $J^{PC} = 0^{++}$   & $m_{\dil'} = \ORD(\La)$ &$F_{\dil'} =  \ORD(\mq^{\frac{\eta_{F_{\dil'}} }{1+\ga_*}} )$
    & $m_{\dil} =  \ORD(\mq^\ym{1})$ & $F_{\dil} =  \ORD(\La)$ 
	\end{tabular}
	\caption{\small Overview of how the important parameters entering the
	explicitly and anomalously broken Ward identities behave in the conformal
	dilaton phase. The scale $\La$ stands for  a generic hadronic scale which in
	QCD is usually referred to as $\La_{\textrm{QCD}}$. 
	The behaviour of $m_{D,\pi}$ 
	and $F_{\dil',\pi'}$ ($\eta_{F_{\dil',\pi'}}/(1+\ga_*) \geq 1$)  under mass-deformation will be discussed in \SEC\ref{sec:m}.
	The
	quantity $\ga_*$ is the mass anomalous dimension at the IR fixed point. }
	\label{tab:summary}
\end{table}
 When expressed in terms
of hadronic quantities, the divergences of the Noether currents are given by
products of decay constants times masses, as shown on the right-hand side of
Eqs.~\eqref{eq:chiWI}, \eqref{eq:dilWI}; their vanishing occurs through
$m_{\pi,\dil} \to 0$ as required by the Goldstone nature of the pions and the
dilaton. The decay constants are the order parameters and do not vanish.
Heuristically one has 
\begin{equation}
  \label{eq:heuristic}
  \textrm{SSB:        }  \qquad Q \state{0} \neq 0 \;, 
  \quad Q = \int d^\dbar x \, J^0 \;, 
\end{equation}
the signal of spontaneous symmetry breaking (SSB), is equivalent to
\eqref{eq:fdecay}.\footnote{There is a subtlety  with this argument in that
the norm of the state created in Eq.~\eqref{eq:heuristic} is proportional to
square root of the spatial volume. This can be seen by considering the $2$-point
function of the currents and integrating over the spatial parts. A careful
treatment for chiral symmetry can be found in Ref.~\cite{Goldstone:1962es}. 
}  
For the non-GBs, which we denote by  $\pi'^{a}$ and $\dil'$, it is just the
opposite, the WIs~\eqref{eq:chiWI}, \eqref{eq:dilWI} are satisfied by a zero
decay constant as the hadronic masses are non-zero. 
For the $D'$ this is a subtle statement in view of the anomalous breaking of scale symmetry but in the end
this is implied by the WI which holds for higher states cf. the remark above.
An overview of the parametric
behaviour in the conformal dilaton phase is given in \TAB\ref{tab:summary} and
the precise mass scalings are discussed in \SEC\ref{sec:def}. Equipped with the
broad picture we summarise the characteristics of the three phases before
getting to the heart of the paper.

\subsection{Overview of the Extended Conformal Window}
\label{sec:overview}

\begin{centering}
\begin{figure}[h!]
\includegraphics[width=1.0\linewidth]{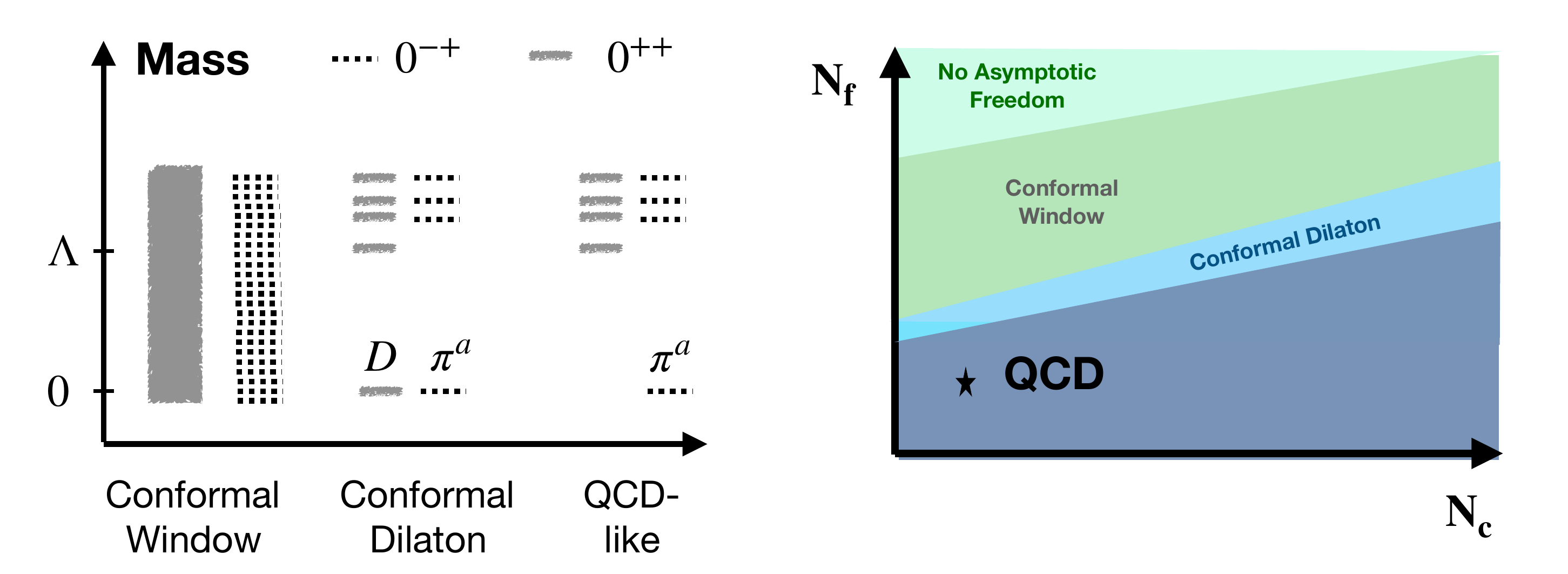}  
	\caption{\small  (left) Schematic spectra of conformal window, conformal
	dilaton and a QCD like theory in the $\mq \to 0$ limit. (right) Qualitative 
	phase diagram of a given matter representation as a function of the number of
	colours and flavours. The boundary of asymptotic freedom is well-established
	and known as the Banks-Zaks region. The boundary with the QCD region is a
	matter of debate. The light-blue conformal dilaton phase is the one discussed
	in this paper. We wish to emphasise that this is just schematic and that the region of this phase 
	could be rather different (should it exist at all).  
	In this paper we discuss its logical possibility and 
	speculate in \SEC\ref{sec:dilaton} that QCD itself could be of this type.}
	\label{fig:overview}
\end{figure}
\end{centering}

Let us summarise the different phases of gauge theories. First, we know from the
Banks-Zaks analysis~\cite{Banks:1981nn} that there is a conformal phase for $N_f
\approx 16$ and $N_c =3$ and probably well below. The range in $N_f$ before
conformal symmetry is (dynamically) broken is known as the conformal window and
its determination is the topic of ongoing efforts of
continuum~\cite{Baikov:2016tgj,Ryttov:2010iz,Ryttov:2016ner,Ryttov:2016hal} and
lattice Monte Carlo
studies~\cite{Hasenfratz:2020ess,Fodor:2019ypi,Fodor:2017gtj,Fodor:2018uih,
Chiu:2018edw,LatticeStrongDynamics:2018hun,DelDebbio:2015byq,Brower:2015owo,LSD:2014nmn,LatKMI:2014xoh,LatKMI:2016xxi}
(cf. \cite{Witzel:2019jbe} for a recent review). In ${\cal N}=1$ supersymmetric
gauge theories this boundary is  known exactly. Below the conformal window chiral
symmetry is spontaneously broken and quark confinement takes place. In
particular this happens in QCD where  $N_c = 3$, $N_f = 3$ (three light
flavours)  and quarks are in the fundamental representation of $SU(N_c)$. What
we are advertising here is that there might be a third phase embedded in the
conformal window where conformal symmetry is spontaneously broken. It would seem
reasonable to assume that this phase lives on the boundary of the conformal
window as sketched in  \FIG\ref{fig:overview}.

The paper is organised as follows. In \SEC\ref{sec:GF}  we define the
gravitational form factors and show how the dilaton restores the dilatational
WI. In \SEC\ref{sec:def} matter mass and finite volume effects are discussed.
Specific search strategies for the conformal dilaton phase with lattice Monte
Carlo simulations are assembled in \SEC\ref{sec:signature} along with a
discussion on whether the  dilaton could be the $f_0(500)$ or the Higgs in QCD or
the electroweak sector. The paper ends with discussion and conclusions in
\SEC\ref{sec:con}. \APPs \ref{app:conv}, \ref{app:12} deal with conventions and
the spin-$1/2$ form factors.

\section{Gravitational Form Factors of Spin-$0$}
\label{sec:GF} 

The gravitational form factors parametrise the matrix elements of the energy
momentum tensor (EMT) between physical states; they can serve as quantum
corrections to external gravitational fields~\cite{Holstein:2006ud}, or as
probes of the nucleon structure~\cite{Leader:2013jra,Hudson:2016gnq}. The
spin-$1/2$ case is discussed in \APP\ref{app:12} and the spin-$1$,
parameterised in~\cite{Holstein:2006ud}, amounts to an interplay between $F_1$
and $F_2$ at zero momentum transfer. Here we focus on the spin-$0$ case since
it illustrates all the important points without unnecessary complications. The
dimensionless gravitational form factors for a generic scalar hadron, denoted by
$\scal$, are defined as follows
\begin{alignat}{2}
  \label{eq:GF}
  & \Tphi_{\mu \nu}(p,p') &\;=\;& \matel{\scal(p')}{T_{\mu\nu}(0)}{\scal(p)} =
  2 {\cal P}_\mu {\cal P}_\nu  \Gone (q^2) +  
  ({{q_\mu q_\nu}} -  
 q^2  \mink_{\mu \nu} ) \frac{m_{\scal}^2}{q^2} \,
  \Gtwo(q^2)  \;,  
\end{alignat}
where $q \equiv p - p'$ is the momentum transfer, $ {\cal P} \equiv \frac{1}{2}
(p + p')$ and $q^\mu \Tphi_{\mu\nu} =0$,  as required by translational
invariance.  This parameterisation is well suited for $m_\scal \neq 0$, 
that is the non-GB sector,  which is the case we aim to examine.
Further note that the limit $q^2 \to 0$ of \eqref{eq:GF} is still well
defined, despite the pole in $\Gtwo$, because for diagonal form factors the
limit implies $q_\mu \to 0$ at the same time. Since the EMT is related to the
momentum, $P_\mu = \int d^\dbar x \, \Tud{0}{\mu}(x)$, by the usual conserved
current procedure, the form factor $\Gone$ must satisfy
\begin{equation}
  \label{eq:Pmu}
  \Gone(0)=1 \;, 
\end{equation}
where we use the conventional state normalisation $\langle \scal(p')|\scal(p)
\rangle = 2 E_p (2 \pi)^3  \de^{(3)}(\veci{p} - \veci{p}') $. 
Note that \eqref{eq:Pmu} holds equally for massless hadrons (e.g. Goldstone bosons) such as the pion or the dilaton. 
 The second
structure is related to the improved energy momentum tensor which renders the
free scalar field conformal in dimension other than two \cite{Callan:1970ze}.
Everything in this section, up to now, was completely general. In the next
section we discuss the conformal IR phases with particular emphasis on the
dilaton case. 
 
\subsection{The Gravitational Form Factors in the Conformal Phase}
\label{sec:GravConformalPhase}

In \SEC\ref{sec:WI} we show that $\Tphiud{\mu}{\mu}(p,p')$, as defined in
\eqref{eq:GF}, vanishes when there is an IR fixed point. This yields one constraint on the form
factors for any spin and in particular for spin-$0$ this results in
\begin{equation}
  \label{eq:spin0con}
  2  m_{\scal}^2 (1- \frac{q^2}{4 m_\scal^2})   \Gone(q^2)   - (\dbar)   m_{\scal}^2 \Gtwo(q^2) = 0  \;.
\end{equation} 
The most straightforward solution is the one of unbroken conformal symmetry
for which  $m_{\scal}^2 =0$ leads to a trivial solution. This is the classic
conformal window scenario. However, there is another possibility for $m_{\scal}^2 \neq 0$: the second
term cancels the first one. In particular this implies 
$\Gtwo(0) = 2 /(d-1)$, taking into account Eqs.~\eqref{eq:spin0con}
and \eqref{eq:Pmu}. And this is where the dilaton pole and spontaneous breaking
of scale symmetry come into play. In summary one has the three 
phases depicted in \FIG\ref{fig:overview}~\footnote{The second
relation can be seen as a cousin of the Goldberger-Treiman relation for the
nucleons. The analogy is not strictly close as there the partially conserved
axial current (PCAC) gives a non-vanishing term on the RHS (which though
vanishes in the limit $\mq \to 0$).  This results in $g_A = 1 + \ORD(m_\pi^2)
\approx 1.23$ (e.g. \cite{Donoghue:1992dd})  and not an exact relation like $\Gone(0)=1$.}$^,$\footnote{
This is the only, straightforward, logical possibility as the $J^{PC} = 0^{++}$ state does not contribute to the $\Gone$ form factor and a composite massless $J^{PC} = 2^{++}$ is forbidden by the Weinberg-Witten theorem \cite{Weinberg:1980kq}.}
\begin{alignat}{5}
  \label{eq:crucial}  
  & \text{Conformal Window:} \quad    & & 
    \Gtwo(q^2) &\;\neq \;& 0 \;, \quad & &  
    \matel{\scal}{\Tud{\mu}{\mu}}{\scal} &\;=\;&  2 m_{\scal}^2 =0 \;, 
     \nonumber \\[0.1cm]
  & \text{Conformal Dilaton:} \quad     &  & \Gtwo(q^2) &\;=\;&
    \frac{2}{\dbar} (1- \frac{q^2}{4 m_\scal^2})  \Gone(q^2)\;,    
    \quad & & \matel{\scal}{\Tud{\mu}{\mu}  }{\scal}   &\;=\;& 0\; ,  \;
    m_{\scal}^2  \neq 0  \;, 
    \nonumber \\[0.1cm]
  & \text{QCD-like:} \quad   & &  \Gtwo(q^2) &\;\neq \;& 0  \;, \quad  & & 
    \matel{\scal}{\Tud{\mu}{\mu}  }{\scal}   &\;=\;& 
  2  m_{\scal}^2  \neq 0  \;.
\end{alignat}
In order to avoid confusion
it seems crucial to state that in this scenario the usual relation $2
m_{\scal}^2 = \matel{\scal}{\Tud{\mu}{\mu}  }{\scal}$ does \emph{not} hold, cf.
above, as   it would either not allow for hadron masses or the dilation WI to be
obeyed.\footnote{This implies that the gluon condensate definition \cite{DelDebbio:2013sta}, 
which departs from this relation, does hold in QCD-like but not in the conformal dilaton phase.} 
 The possibility of such a scenario was mentioned prior to the discovery of 
the trace anomaly 
\cite{Gell-Mann:1969rik} but not worked out, for example in terms of hadronic
parameters.  The doing thereof is the topic of the next section.

\subsection{Verification of Dilatation Ward Identity at $q^2=0$ via the
LSZ formalism}
\label{sec:LSZDilWI}

It is advantageous to represent  the form factor $\Gtwo$ in terms of a subtracted dispersion relation 
\begin{equation}
\Gtwo(q^2) = \Gtwo(0) + \frac{q^2}{\pi} \int_{0+}^\infty \frac{ds \, \textrm{Im}[\Gtwo(s)]}{s(s-q^2-i0)} \;,
\end{equation}
where $0^+$ indicates that the single dilaton has been removed from the integral. From \eqref{eq:spin0con} we infer the low energy theorem $\Gtwo(0) = 2/(d-1)$ which  we are able 
to verify explicitly, using the LSZ formalism (e.g.~\cite{Donoghue:1992dd,Weinberg:1995mt,Duncan:2012aja}), as this point corresponds to the on-shell process  $\scal
\to \scal \dil$. 
The effective
Lagrangian for the $\scal \to \scal \dil$ process is 
$ {\cal L}^{\textrm{eff}} = g_{\scal \scal \dil } \frac{1}{2} \scal^2 \dil$,
which yields the amplitude
\begin{equation}
  \label{eq:OS}
  \vev{\dil \scal | \scal } =  i (2\pi)^d \de\left(\sum p_i\right)\, 
  g_{\scal \scal \dil } \, .
\end{equation}
To achieve our goal two steps are needed. First we need to determine $g_{\scal
\scal \dil }$ in terms of other parameters and then we apply the LSZ reduction
to extract $\vev{\dil \scal | \scal }$ and match to \eqref{eq:OS}. The $g_{\scal
\scal \dil}$ coupling can be determined by writing an effective Lagrangian for
the dilaton, where the field $e^{D(x)/F_D}$ plays the role of a conformal
compensator, see e.g.~\cite{Coleman:1985rnk}. Namely, terms in the Lagrangian
which scale like $\sqrt{-g}{\cal L} \to  e^{- n \al } \sqrt{-g}{\cal L} $ under
dilatations $g_{\mu\nu} \to e^{-2 \al} g_{\mu\nu}$ can be made  invariant by
adding a prefactor $e^{-n \dil/F_\dil}$, where $\dil \to \dil -  \al F_{\dil}$
under scale transformations.\footnote{Here $g_{\mu \nu} = \eta_{\mu\nu}$ and 
$g = \det(g_{\mu\nu})$
denotes the determinant and therefore $\sqrt{-g} \to e^{-d \al} \sqrt{-g}$ under Weyl transformation.
  Note that our sign convention of $F_{\dil} $ is opposite as compated  
to some of the literature, e.g. \cite{Cata:2018wzl}, in order to preserve the analogy with the pion 
decay constant \eqref{eq:fdecay}. Hence the change of sign in formulae with $F_{\dil}$ as compared 
to these works.
In the case where the transformation parameter is
chosen to be a local function one often refers to these transformations as Weyl
scaling. The term below is Weyl invariant.}  Applied to the mass terms this
gives the following appropriate  effective Lagrangian ($\scal \to e^{\al} \scal \Rightarrow 
 {\cal L}^{\textrm{eff}}  \to e^{4 \al}  {\cal L}^{\textrm{eff}} $) 
\begin{equation}
  \label{eq:gSSD}
  {\cal L}^{\textrm{eff}} = - e^{- 2\dil/F_{\dil}} \,    \frac{1}{2} m_\scal^2 \scal^2 
  \quad \Rightarrow \quad g_{\scal \scal \dil } = 
   \frac{2 m_\scal^2}{F_{\dil}} \;.
\end{equation}
Second the  matrix element in~\eqref{eq:OS} can be obtained in another way,
directly from the form factor~\eqref{eq:GF}, by using the EMT as an
interpolating operator of the dilaton.  
We are interested in the dilaton appearing in the $ ({{q_\mu q_\nu}} -  
 q^2  \mink_{\mu \nu} )$-structure for which it is straightforward to write down a projector $P_2$, such that  $P_2^{\mu\nu}  ({{q_\mu q_\nu}} -  
 q^2  \mink_{\mu \nu} ) =1$ and $P_2^{\mu\nu}   P_\mu P_\nu =0$.
The on-shell matrix element then follows from 
\begin{alignat}{2}
\label{eq:analogy1}
  & \vev{\dil \scal | \scal } &\;=\;& 
  \lim_{q^2 \to  0} (-i) \frac{q^2}{Z_{\dil}} 
  \int d^d x e^{i q\cdot x} P_2^{\mu\nu}  \Tphi_{\mu\nu}(p,p',x)   \nonumber \\[0.1cm]
  & &\;=\;&  \lim_{q^2 \to  0} (-i) \frac{q^2}{Z_{\dil}}  \frac{m_{\scal}^2}{q^2}  \Gtwo(q^2) (2\pi)^d  \de\left(\sum p_i\right)  \;,
\end{alignat}
 where $Z_{\dil} =  - F_{\dil}/(\dbar)$ is, by  footnote~\ref{foot:fD}, the corresponding LSZ factor.  
Identifying
the two equations one gets, using~\eqref{eq:gSSD}
\begin{equation}
  \label{eq:new}
  \lim_{q^2 \to 0}  \Gtwo(q^2) 
  = - \frac{{g_{\scal \scal \dil } Z_{\dil}} }{m_{\scal}^2} = 
   \frac{2}{d-1} \;,
\end{equation}
which satisfies \eqref{eq:crucial} when $\Gone(0)=1$ is taken into account. This matches~\eqref{eq:spin0con} in the $q^2 \to 0$ limit  and thus shows
that a dilaton phase  seems a logical possibility indeed. The interplay
of the dilaton residue and the vanishing of the trace of the EMT is an
encouraging result.

\section{Perturbations of the Conformal Dilaton Phase}
\label{sec:def}

\subsection{Quark Mass-Deformation}
\label{sec:m}

We turn now to the question of how the hadronic quantities change when the quark
mass is turned on. At a scale $\sqrt{q^2} \ll \La$, introduced in \TAB\ref{tab:summary}, all  states except
the dilaton and the pion decouple from the spectrum and we essentially have a conformal
theory with a dilaton and pions. This situation is similar to the
mass-deformed conformal window scenario  extensively discussed in our previous
papers~\cite{DelDebbio:2010ze,DelDebbio:2010jy,DelDebbio:2013qta}, provided that
$\mq \ll \La$ (as otherwise the quarks would decouple). 
The result
that is sufficient for this section is that a matrix element of an operator
${\cal O}$, of scaling dimension $\Delta_{\cal O} = d_{\cal O}+ \ga_{\cal O}$,
between physical states $\phi_{1,2}$ in the vicinity of the fixed point behaves like \footnote{In our previous work this was shown to hold on the lowest state in each channel, except for the masses where it was shown in generality \cite{DelDebbio:2010jy}. However, our arguments at the end of  \SEC\ref{sec:WI}  shows that it holds for all states.}
\begin{equation}
  \label{eq:hyper}
  \matel{\phi_2}{{\cal O}}{\phi_1}  =\mq^{\ym{\eta}} \;, 
  \quad \eta =  \Delta_{\cal O} + d_{\phi_1} + d_{\phi_2} \;,
\end{equation}
where we have assumed zero momentum transfer ($p=p'$) for the time being.
Above $d_{\cal O}$ and $\ga_{\cal O}= - \frac{d}{d\ln \mu} \ln {\cal O}$ stand for the
engineering and anomalous dimensions respectively.
The relation \eqref{eq:hyper} has limited applicability in our case because of the presence 
of the additional scale $\La$, a point we will return to in the next section.
We can only apply it to the dilaton and the pion mass. 
 Starting from  \eqref{eq:hyper}  one can
obtain a differential equation, using the trace anomaly, which leads
to~\cite{DelDebbio:2010jy}
\begin{equation}
  \label{eq:MassScaling2010}
  m^2_{\dil} \propto \mq^{\ym{2}}   \;,\quad m^2_{\pi} \propto \mq^{\ym{2}}  \;.
  \end{equation}
  Alternatively this result can be obtained following other techniques \cite{DelDebbio:2010ze}
  which correspond to setting $\La = 0$ in \SEC\ref{sec:ScalingDynamical}.

It is also of interest 
to investigate the scaling of $F_{\dil',\pi'}$ which can be done by using 
the dilaton WI \eqref{eq:dilWI} applied to $\dil',\pi'$
\begin{alignat}{2}
&  {F_{\dil'}} m_{\dil'}^2  &\;=\;&  \matel{0}{  \mq(1\pl \ga_*) 
    \bar{q}q }{\dil'(q)}  \propto \mq^{\frac{\eta_{\comA{F_{\dil'}} m_{\dil'}^2}}{1+\ga_*}} \;, \quad  
  {  \small{ \frac{\eta_{\comA{F_{\dil'}} m_{\dil'}^2} }{1+\ga_*} \geq 1}  } \;, \nonumber \\[0.1cm]
  &  {F_{\pi'}} m_{\pi'}^2  &\;=\;&  \matel{0}{  \mq(1\pl \ga_*) 
    \bar{q}q }{\pi'(q)}  \propto \mq^{\frac{\eta_{\comA{F_{\pi'}}  m_{\pi'}^2}}{1+\ga_*}} \;, \quad  
  {  \small{ \frac{\eta_{\comA{F_{\pi'}} m_{\pi'}^2} }{1+\ga_*} \geq 1}  } \;,
\end{alignat}
where the matrix element proportional to the $\be$-function has been neglected, as it is subleading 
for $\mq \approx 0$.  
The statement 
$\eta_{F_{\dil',\pi'} m_{\dil',\pi'}^2}/(1+\ga_*) \geq 1$ then follows from the assumption that 
the matrix element  \comA{$\matel{0}{ \bar{q}q }{\dil',\pi'(q)} $} is finite 
for $\mq \to 0$. Since $m_{\dil'} = \ORD(\La)$ it then follows that 
\begin{equation}
{F_{\dil',\pi'}} \propto \mq^{\frac{\eta_{F_{\dil',\pi'}}}{1+\ga_*}} \;, \quad
  {  \small{ \frac{\eta_{\comA{F_{\dil',\pi'}}}}{1+\ga_*} \geq 1  }  }\;.
 \end{equation}
These observations are interesting per se and complete \TAB\ref{tab:summary} but
we would like to understand how~\eqref{eq:spin0con} is altered. 
We may use the same WI as above but applied to a diagonal matrix element  and conclude  
 \begin{equation}
  \label{eq:LHSm}
  \Tphiud{\mu}{\mu} (0)  =  \mq (1+ \ga_*)\matel{\scal}{\bar qq}{\scal}\propto  \mq^{\frac{\eta_{T_\phi}}{1+\ga_*}} \;, \quad  {  \small{ \frac{\eta_{T_\phi}}{1+\ga_*} \geq 1  }  } \;.
 \end{equation}
 The first scaling follows from the hyperscaling relation  \eqref{eq:hyper} and the second one, once more, 
 from the assumption that the matrix element  $\matel{\scal}{\bar qq}{\scal}$  is finite as $\mq \to 0$.  
  The correction to the form factor constraint \eqref{eq:spin0con} then follows from the correction to 
 the  on-shell coupling~\eqref{eq:gSSD} and leads to 
\begin{equation}
  \label{eq:G20m}
  \Gtwo(0)= 
  \frac{2}{\dbar}\left[1 + 
  \ORD \left(\mq^{\frac{\eta_{T_\phi}}{1+\ga_*}}\right) \right]\;,
\end{equation}
since $\Gone(0) =1$ in general.
Hence the scaling correction will come from the second term in~\eqref{eq:G20m} 
so that the RHS can match~\eqref{eq:LHSm}. 

An interesting question is how this changes when the momentum transfer is non-zero.  
We may assess this question by expanding in $q^2$
\begin{equation}
 \Tphiud{\mu}{\mu} (q^2) =  \Tphiud{\mu}{\mu} (0) + q^2 \frac{d}{d q^2}   \Tphiud{\mu}{\mu} (0) + 
 \frac{q^4}{2}   \left( \frac{d}{d q^2} \right)^2   \Tphiud{\mu}{\mu} (0) + \ORD(q^6) \;,
\end{equation}
and demanding that the expansion converges which amounts to determine the scaling of the derivatives. 
First we note, cf.  \SEC\ref{sec:WI} for more details, that $\Tphiud{\mu}{\mu} (q^2) =0$ for $\mq \to 0$ 
and  thus we may apply the RG analysis  in \SEC 3 of our previous work \cite{DelDebbio:2013qta} 
as applied to the pion form factor.  We  infer that 
\begin{equation}
\left( \frac{d}{d q^2} \right)^n  \Tphiud{\mu}{\mu} (0)  \propto  \left( \frac{\mq}{\La} \right)^{ \frac{ \eta_{T_\phi}}{1+\ga_*}} 
\left( \frac{1}{\La_m^2} \right) ^{n} \;,    \quad  \La_m  \equiv \mq^{\frac{1}{1+\ga_*}}  \La^{\frac{\ga_*}{1+\ga_*}} \;,
\end{equation} 
where the first factor is just the previous result in \eqref{eq:LHSm} and $\La_m$ sets the new scale. 
Note that the relative coefficients, unlike $\eta_{T_\phi}$ itself,  of the form factor derivatives follows the straightforward hyperscaling law 
as they  are not affected by the dynamical scale $\La$ to be assessed in the next section.  
Hence  $\Tphiud{\mu}{\mu} (q^2) $ ceases to be close to  $\Tphiud{\mu}{\mu} (0)  \propto (\mq/\La) ^{\eta_{T_\phi}/(1+\ga_*)}$ for momentum transfers $q^2 \gg\La_m^2$.
In some sense the scale $\La_m$ defines the deep IR for which the TEMT reveals its IR fixed-point 
in the presence of an explicit quark mass $\mq$.

\subsection{Scaling in the Presence of a Dynamical Scale $\La$}
\label{sec:ScalingDynamical}

Let us now revisit the RG scaling for field correlators in the case where scale
invariance is spontaneously broken. We closely follow the derivations in our
previous studies~\cite{DelDebbio:2010jy}, allowing for the dependence on an
extra scale $\Lambda$ that is dynamically generated as a result of the
spontaneous breaking.  If it was not for the scale $\La$ one would directly conclude 
that $\eta_{T_\phi} = 2$ and $\eta_{F_{\dil'} m_{\dil'}^2} = 3$. In this section we  shall see 
why this conclusion does not hold in the presence of the dynamical scale $\La$.

We  consider both $2$-point and $3$-point functions
in Euclidean space, which are defined respectively as 
\begin{equation}
  \label{eq:TwoPtDef}
  C_{{\cal O}\Phi}\left(t, \mathbf{p}, \Lambda; 
  \delta g, \hat{m}_q, L^{-1}, \mu\right) =
  \frac{1}{ V} \int d^{\dbar}x\, e^{-i \mathbf{p} \cdot \mathbf{x}}\, 
  \langle 0 | {\cal O}(t, \mathbf{x})\Phi(0)^\dagger | 0\rangle \;,
\end{equation}
where $p^0= i E_p = i \sqrt{\mathbf{p}^2+m_\phi^2}$, and 
\begin{equation}
  \label{eq:ThreePtDef}
  \begin{split}
    C_{{\cal O}\Phi\Phi}&\left(T, t, \mathbf{p}, \mathbf{p}', \Lambda; 
    \delta g, \hat{m}_q, L^{-1}, \mu\right) = \\
    & \quad \quad =\frac{1}{V^2} \int d^{\dbar}x\, d^{d-1}y\, 
    e^{-i ( \mathbf{p}\cdot \mathbf{x}+\mathbf{p}'\cdot \mathbf{y})}\, 
    \langle 0 | {\cal O}(t, \mathbf{x}) \Phi(T,\veci{y})  
    \Phi(0)^\dagger | 0\rangle\;,
  \end{split}
\end{equation}
with $\Phi$ an interpolating field for the particle $\scal$. The theory is
assumed to be defined in a finite volume of linear size $L$ and at a scale
$\mu$, in the neighbourhood of a RG fixed point,  located at $\delta g=\hat{m}_q=0$. The couplings $\delta g$ and $\hat{m}_q$ are
both dimensionless. If necessary, dimensionful couplings are rescaled by the
appropriate powers of the scale $\mu$. The spatial volume is $V=L^{d-1}$. 

Using unitarity and usual RG scaling arguments -- see e.g. our previous
publications \cite{DelDebbio:2010ze,DelDebbio:2010jy,DelDebbio:2013qta} for
details -- we obtain
\begin{alignat}{2}
  & C_{{\cal O}\Phi} \left(t, \mathbf{p}, \Lambda; 
    \delta g, \hat{m}_q, L^{-1}, \mu\right) &\;=\;&     
    \frac{e^{-E_p t}}{ 2 E_p \, V }\, \langle 0 | {\cal O}(0) | \scal (\mathbf{p})
    \rangle\, \langle \scal (\mathbf{p}) | \Phi(0)^\dagger | 0 \rangle +
    \ldots\,  \nonumber \\
  \label{eq:TwoPtScaling}
  & &\;=\;& b^{-\Delta_{\cal O} - \Delta_\Phi}\, C_{{\cal O}\Phi} \left(b^{-1}
    t, b \mathbf{p}, b\Lambda; b^{y_g}\delta g, b^{y_m}\hat{m}_q, b L^{-1},
    \mu\right)\, ,
\end{alignat}
where we 
have used
 the identification $ \comA{(2\pi)^{(d-1)}} \delta^{(d-1)}(0) \leftrightarrow V$ and
 $\Delta_{\cal O}$ and $\Delta_\Phi$  are the scaling dimensions of the
operators ${\cal O}$ and $\Phi$.  The quantities  $y_g$ and $ y_m \equiv
1+\ga_*$ are the critical exponents that characterise the running of
the couplings determined by the linearised RG equations in the vicinity of the
fixed point. $\scal(\mathbf{p})$ is the lightest state in the spectrum with the
same quantum numbers as $\Phi(x)$ and energy $E_p=\sqrt{\mathbf{p}^2+m_\phi^2}$. The
ellipses represent the contributions from excited states in the spectrum, which
are exponentially suppressed. 

The scaling formula for the $2$-point function can be used as usual to derive 
the scaling of the masses of the hadronic states. Setting $\mathbf{p}=0$, and 
$b^{y_m}\hat{m}_q=1$ yields
\begin{equation}
  \label{eq:DecouplingScaling}
  \begin{split}
  C_{{\cal O}\Phi} &\left(t, 0, \Lambda; 
    \delta g, \hat{m}_q, L^{-1}, \mu\right) = 
   \hat{m}_q^{\frac{\De_{\cal O} +\De_\Phi}{y_m}} \, 
   C_{{\cal O}\Phi} (\hat{m}_q^{1/y_m} t, 0, \hat{m}_q^{-1/y_m}\Lambda; 0,1,\mu) \,+ \\[0.1cm]
    & \qquad \qquad\qquad \qquad\qquad \qquad \qquad \qquad \ORD\left(\mq^{-y_g/y_m} \delta g\right)\, .    
  \end{split}
\end{equation}
We may parameterise the large-$t$ behaviour as 
\begin{equation} 
 C_{{\cal O}\Phi} \left(t, 0, \Lambda; 
    \delta g, \hat{m}_q, L^{-1}, \mu\right)  \to  \hat{m}_q^{\frac{\De_{\cal O} +\De_\Phi}{y_m}}  e^{- \left(
    \hat{m}_q^{1/y_m} \mathcal{F}(\hat{m}_q^{-1/y_m} \Lambda) \,  \La  t \right) } f( \hat{m}_q^{-1/y_m} \Lambda),\mu) \;,
\end{equation}
where both functions, ${\cal F}$ and $f$, can and will overturn the hyperscaling
behaviour found in \eqref{eq:hyper} for masses and matrix elements. Specifically
we may read off the behaviour of the $\phi$-mass
\begin{equation}
  \label{eq:MassScalingDecoupling}
  m_{\phi}  \propto  \hat{m}_q^{1/y_m}\, \mathcal{F}(\hat{m}_q^{-1/y_m} \Lambda)\, .
\end{equation}
We are interested in the scaling of the masses as the fermion mass $\mq \to 0$,
which corresponds to $\mq^{-1/y_m} \Lambda \to \infty$. We can then distinguish two
different regimes
\begin{alignat}{3}
  \lim_{x\to\infty} \mathcal{F}(x) = 
   \left\{\begin{array}{lll}   \kappa   &\quad \Rightarrow &
    \quad   m_{\phi} =  \ORD\left(\mq^{\ym{1}}\right)   \\ 
     \kappa x   & \quad \Rightarrow &
    \quad   m_{\phi} =  \ORD\left( \La \right)  
    \end{array} \right.
  \;,  
\end{alignat}
where $\kappa$ is a constant and the first case is an alternative derivation 
of the mass scaling quoted earlier. 
The first regime corresponds to the conformal
scaling already discussed in our previous study~\cite{DelDebbio:2010jy}.
Interestingly, the second regime yields the scaling with $\Lambda$ that is 
expected in the theory with spontaneously broken symmetry and a dilaton.
We further note that, using arguments about the finiteness of matrix elements in
the $\mq \to 0$ limit (as done in \SEC\ref{sec:m}), it may be possible to make
further statements about the function $f(x,\mu)$ as $x \to \infty$. We refrain
from doing so as it does not add anything to the key messages of this paper.

\subsection{Dilaton Ward Identity in the Vicinity of the IR Fixed Point}
\label{sec:WI}

A similar analysis for the $3$-point function allows us to derive a crucial  result for the WI in the neighbourhood of a fixed point.
Once again we start from the RG equation, 
\begin{align}
  C_{{\cal O}\Phi\Phi}& \left(T, t, \mathbf{p}, \mathbf{p}', \Lambda; 
    \delta g, \hat{m}_q, L^{-1}, \mu\right) = 
    \nonumber \\
    & = \frac{e^{-E_p(T-t)} e^{-E_{p'}t}}{  2 E_p\, 2 E_{p'} \, V^2}\, 
      \langle 0 | \Phi(0) | \scal(\mathbf{p}) \rangle \, 
      \langle \scal(\mathbf{p}) | {\cal O}(0) | \scal(\mathbf{p}') \rangle \, 
      \langle \scal(\mathbf{p}') | \Phi(0)^\dagger | 0\rangle 
      + \ldots \nonumber \\
    \label{eq:ThreePtScaling}
    & = b^{-\Delta_{\cal O} - 2 \Delta_\Phi} 
    C_{{\cal O}\Phi\Phi}
    \left(b^{-1} T, b^{-1} t, b \mathbf{p}, b \mathbf{p}', b \Lambda; 
    b^{y_g} \delta g, b^{y_m} \hat{m}_q, b L^{-1}, \mu\right) \, .
\end{align}
Combining these expressions, we obtain the matrix elements from taking the
large-time limits of correlators. In particular, we have
\begin{align}
  \langle 0 | {\cal O}(0) | \scal(\mathbf{p}) \rangle 
    &= K_2 \lim_{b\to\infty}\, e^{E_P bt}\, 
      C_{{\cal O}\Phi}(bt, \mathbf{p}, \Lambda; 
      \delta g, \hat{m}_q, L^{-1}, \mu) \nonumber \\
    \label{eq:MELimitFour}
    &= K_2 \lim_{b\to\infty}\, e^{E_P bt}\, b^{-\Delta_{\cal O}-\Delta_\Phi}\,
      C_{{\cal O}\Phi}\left(t, b \mathbf{p}, b \Lambda; 
      b^{y_g}\delta g, b^{y_m}\hat{m}_q, b L^{-1}, \mu\right)\, ,
\end{align}
and similarly
\begin{align}
  \langle &\scal(\mathbf{p}) | {\cal O}(0) | \scal(\mathbf{p}') \rangle 
   = \nonumber \\ 
  &= K_3\, 
    \lim_{b\to\infty} e^{E_p b(T-t)}\, e^{E_{p'} bt}\, 
    C_{{\cal O}\Phi\Phi}\left(bT, bt, \mathbf{p}, \mathbf{p}',\Lambda; 
    \delta g, \hat{m}_q, L^{-1}, \mu\right) \nonumber \\
  \label{eq:MELimitTwo}
  &= K_3 \lim_{b\to\infty} e^{E_p b(T-t)}\, e^{E_{p'} bt}\, 
    b^{-\Delta_{\cal O}-2\Delta_\Phi}\, C_{{\cal O}\Phi\Phi}
    \left(T, t, b \mathbf{p}, b \mathbf{p}', b\Lambda; 
    b^{y_g} \delta g, b^{y_m} \hat{m}_q, b L^{-1}, \mu\right) \, ,
\end{align}
where 
\begin{align}
  \label{eq:K2Def}
  K_2 &= \left(\frac{ 
  \langle \scal(\mathbf{p}) | \Phi(0)^\dagger | 0\rangle}
  {  2 E_p \, V }\right)^{-1}\;, \quad \\
  \label{eq:K3Def}
  K_3 &= \left(\frac{\langle 0 | \Phi(0) | \scal(\mathbf{p}) \rangle 
  \langle \scal(\mathbf{p}') | \Phi(0)^\dagger | 0\rangle}
  { 2 E_p\, 2 E_{p'} \, V^2  }\right)^{-1}\,.
\end{align}

Eqs.~\eqref{eq:MELimitFour} and~\eqref{eq:MELimitTwo} are the master formulae
needed in order to understand the IR behaviour of the dilatation WI and the
scaling of finite-volume effects.  From these formulae one infers that evaluating the correlation functions
at infinite time separation is the same as evaluating them at finite time with other dimensionful parameters 
appropriately rescaled.
Now, taking the infinite-volume limit first, we are
able to show that the on-shell WIs are insensitive to the anomalous breaking in
the presence of an IR fixed point, provided that the explicit breaking of scale
invariance due to the mass is tuned to zero. In order to prove this statement,
we are going to consider the anomalous contribution in
Eq.~\eqref{eq:DivNoetherCurrTwo} due to the gauge field, for $\hat{m}_q=0$, namely
\begin{equation}
  \label{eq:AnomalousOnShell}
  \matel{0}{ \frac{\be}{2 g} G^2}{\dil(q)}\, .
\end{equation}
This matrix element can be obtained from the large-$t$ behaviour of the
correlator $C_{\Tud{\mu}{\mu} \Phi_D}(t)$ where ${\cal O}= \Tud{\mu}{\mu} = \frac{\be}{2 g} G^2$ and
$\Phi_D$ is a generic interpolating operator that has an overlap with the dilaton
field  but not the vacuum (e.g. $\Phi_D \to \Phi_D - \vev{\Phi_D} | 0 \rangle \langle 0 |$ is a realisation thereof).  Starting from the infinite-volume theory and setting $L^{-1}=0$, we
obtain from Eq.~\eqref{eq:MELimitFour}
\begin{align}
  \label{eq:LargeTIRFixedPoint}
  \matel{0}{ \frac{\be}{2 g} G^2}{\dil(q)} \propto
  \lim_{b\to\infty} e^{E_q bt} b^{-\Delta_{\cal O}-\Delta_\Phi}\, 
  C_{\Tud{\mu}{\mu}\Phi_D}(t, b \mathbf{q}, b \Lambda; 
  b^{y_g} \delta g, b^{y_m} \hat{m}_q, 0, \mu)\, .
\end{align}
Note that we need to keep a finite, non-vanishing mass, or a non-vanishing
spatial momentum, in order to guarantee the exponential fall-off of the
correlator. We see from the expression above that in the neighbourhood of an IR
fixed point, the coupling $g$ is irrelevant (that is the critical exponent is negative $y_g<0$).
Assuming that the matrix element of $G^2$ does not diverge in the IR, the
matrix element  therefore vanishes 
\begin{equation}
  \label{eq:IrrelevantVanishes}
 \matel{0}{ \frac{\be}{2 g} G^2}{\dil(q)} \propto   \left.\frac{\beta(g)}{g}\right|_{b^{y_g} \delta g} 
  \propto \frac{b^{y_g} \delta g}{g^*}\, \left[1 - \frac{\delta g}{g^*}\right]
  \to 0 \, .
\end{equation}
Above $g^*$ is the value of the coupling at the IR fixed point.
Eq.~\eqref{eq:IrrelevantVanishes} shows that the anomalous breaking does not
contribute to the WI between the vacuum and the dilaton state.
The only assumption needed is that  the gluonic matrix element remains finite when $\mu \to 0$. The order of the limits is
relevant here: the mass of the matter fields guarantees that the dilaton is
massive and its correlators decay exponentially, then in the large-time limit
the contribution from the running of the gauge coupling vanishes. A similar
argument applied to the 3-point functions shows that the matrix element of the
anomalous breaking term between two one-particle states also vanishes in the
presence of an IR fixed point.  Note that these statements are true not only for the lowest state 
as one may choose an interpolating operator which has no overlap with the lowest state. This is 
particularly clear in the finite volume formulation where the fields can be represented in form of a discrete spectral sum.
In summary we thus have that \footnote{\label{foot:G2} Since the trace of the EMT is a RG invariant this implies $\matel{0}{ G^2(\mu) }{0}=0$, $\matel{0}{ G^2(\mu) }{\phi_1(p)}=0$  and 
$\matel{\phi_2(p')}{ G^2(\mu) }{\phi_1(p)}=0$  for any scale $\mu > 0$.  By continuity it is then also implied for $\mu = 0$. This implies that  the gluon condensate is not the operator that breaks the dilatation symmetry spontaneously.}
\begin{equation}
\label{eq:imp}
\matel{0}{\Tud{\mu}{\mu} }{0}  \stackrel{\textrm{sym}}{\to } 0 \;, \quad 
 \matel{0}{\Tud{\mu}{\mu} }{\phi_1(q)}  \stackrel{\textrm{sym}}{\to } 0 \;, \quad 
  \matel{\phi_2(p')}{\Tud{\mu}{\mu}}{\phi_1(p)}  \stackrel{\textrm{sym}}{\to } 0 \; \;,
\end{equation}
in the $\mq \to 0$ limit where $\phi_{1,2}$ are any physical states 
and the equation also holds for the vacuum expectation value if $\Phi$ is chosen to have overlap with the 
vacuum.
Colloquially speaking, the physical matrix elements ``see" the TEMT at large distances and since there 
is an IR fixed point this means effectively that $ \Tud{\mu}{\mu}  \to 0$ between physical states.
This is an important result of our paper and in agreement with statements found in \cite{Cata:2018wzl}.
In order to delimit this result we stress that  correlation functions  with 
$ \Tud{\mu}{\mu}$-insertions are generically non-vanishing. For example in the context of the flow theorems
they constitute the main observables \cite{Komargodski:2011vj,Shore:2016xor,Prochazka:2017pfa}. 
It seems worthwhile to clarify that  the TEMT does not need to vanish 
on quark and gluon external states since they are not (asymptotic) physical states 
even in the absence of confinement. 
 This is the case since quarks and gluons can  emit  soft coloured gluons and thus  colour is not a good asymptotic  quantum number.  Another aspect, that has the same root, is that quark and gluons correlation functions can have unphysical singularities  on the first sheet. 

\subsection{Finite Volume Scaling}
\label{sec:FiniteV}

Finally, by keeping the size of the system $L$ finite, the solutions
of the RG equations presented above allow us to quantify the scaling of the
correlators in finite (but sufficiently large) volumes. Choosing a reference
scale $L_0$ and setting $b=L/L_0$, we obtain for the $2$-point function \eqref{eq:TwoPtDef}
\begin{equation}
  \label{eq:TwoPtFiniteVol}
  \begin{split}
    \frac{e^{-E_p t}}{ 2 E_p \, V }\, 
      &\langle 0 | {\cal O}(0) | \scal (\mathbf{p})
      \rangle\, \langle \scal (\mathbf{p}) | \Phi(0)^\dagger | 0 \rangle +
      \ldots\,  \\
      =& \left(\frac{L}{L_0}\right)^{-\Delta_{\cal O}-\Delta_\Phi} \times \\
      & \times C_{{\cal O}\Phi}\left(
        \left(\frac{L}{L_0}\right)^{-1} t, 
        \left(\frac{L}{L_0}\right) \mathbf{p},
        \left(\frac{L}{L_0}\right) \Lambda; 
        \left(\frac{L}{L_0}\right)^{y_g} \delta g, 
        \left(\frac{L}{L_0}\right)^{y_m} \hat{m}_q, 
        L_0,\mu
      \right)\, .
  \end{split}
\end{equation}
This equation allows us to derive the scaling of the energy and of the matrix
elements with the size of the lattice; ignoring the contribution of the
irrelevant coupling, we obtain
\begin{align}
  \label{eq:MassVolScaling}
  & E_p L = f\left(L^{y_m} \hat{m}_q, L \Lambda\right)\, , \\
  & \langle 0 | {\cal O}(0) | \scal (\mathbf{p}) \rangle 
  \propto \left(\frac{L}{L_0}\right)^{-\Delta_{\cal O}}\, , 
\end{align}
as already discussed in our previous studies~\cite{DelDebbio:2010ze}. It is
interesting to emphasise that in a finite volume the anomalous contribution to
the WI from the irrelevant coupling is proportional to
$\left(\frac{L}{L_0}\right)^{y_g}$. Hence, the finite volume explicitly
breaks the scale symmetry by acting as an IR regulator and this is reflected in
the WI for ${\cal O} = \Tud{\mu}{\mu}$. Once again, because $y_g<0$, the
breaking term vanishes when $L\to\infty$, which is consistent with the fact that
SSB cannot occur in a finite volume, as otherwise tunnelling rates prohibit SSB 
\cite{Weinberg:1996kr}.

\section{Conformal Dilaton Signatures}
\label{sec:signature}

In \SEC\ref{sec:lattice} we discuss concretely how the conformal dilaton phase
can be searched for on the lattice and in \SEC\ref{sec:dilaton} we comment on
the ideas that the $f_0(500)$ in QCD and the Higgs could be  dilatons from the
perspective of this paper and the newly obtained scaling formula for its mass. 

\subsection{Lattice Monte Carlo Simulations}
\label{sec:lattice}

In order to discriminate a conformal dilaton phase from the QCD or unbroken
conformal phase we propose the following two strategies.
\begin{itemize}
  \item To test the subtle cancellation in the trace anomaly between the form
  factors  $\Gone(0)=1$ and 
  $\Gtwo(0) = 2 /(d-1)(1+ \ORD(\mq^{2/(1+\ga_*)}))$ (or more generally
  \eqref{eq:spin0con} with $\mq$-corrections), which allows the hadrons to carry mass and the trace
  anomaly to vanish (in the IR). As stated in the main text $\Gone(0)=1$ holds
  irrespective of the phase and it is~\eqref{eq:G20m} that provides the test.
  Neither in QCD nor in the  unbroken conformal phase does $\Gtwo$ exhibit a
  pole at $m_{\dil} = \ORD(\mq^{1/(1+\ga_*)})$ as there is no dilaton.   
  The same applies for the spin-$1/2$ form factors, presented in \APP\ref{app:12},
  for which $g_1(0) =1$ and 
  $g_3(0) = 4 m_{\bary}^2 /(d-1)(1+ \ORD(\mq^{2/(1+\ga_*)})$ (or more generally
  \eqref{eq:spin12con} with $\mq$-corrections) provide the test.

  \item Out of the peculiar properties of the dilaton phase summarised in
  \TAB\ref{tab:summary}, $m_{\dil} =  \ORD(\mq^{1/(1+\ga_*)})$ seems the most
  promising to test. The signal of the conformal dilaton phase is then that for
  $J^{PC} = 0^{++}$
  \begin{equation}
    \label{eq:mdisc}
    m_{\dil} =  \ORD\left(\mq^\ym{1}\right) \;, \quad m_{\pi^a} =  
   \comA{\ORD\left(\mq^\ym{1}\right)}  \;,
    \quad m_{\textrm{other}} =  \ORD(\La) \;. 
  \end{equation}
  This contrasts  QCD where all masses, except the pions, are $\ORD(\La)$ and
  the unbroken conformal window where all masses scale like $m_{\textrm{all}} =
  \ORD(\mq^{1/(1+\ga_*)}) $.
\end{itemize}
We would think that the first test is more spectacular but it might be more
costly as the form factor necessitates 3-point functions whereas masses (and
decay constants) can be extracted from 2-point functions. 
 
\subsection{The Higgs and the $f_0(500)$ as Pseudo-Dilatons}
\label{sec:dilaton}
 
In this section we briefly discuss whether the Higgs or the $f_0(500)$ are
(pseudo)-dilatons in the electroweak and the QCD sector respectively.  Our work
is distinct from other approaches in the scaling formula for the dilaton
\eqref{eq:mdisc} and we mainly focus on this aspect. We would like to
stress that the $\matel{0}{\Tud{\mu}{\mu} }{0}  =0$ for $\mq=0$, in the context
of an IR fixed point,  is of course of interest to the cosmological constant
problem. Moreover, if masses are added for quarks and techniquarks,  they would
decouple in the deep IR. The question of IR conformality is then shifted to the
pure Yang-Mills sector.  Whereas lattice studies indicate that pure Yang-Mills
is confining, it is, to the best of our knowledge, an open question whether these
theories show an IR fixed point not. If this was the case then the Higgs sector
and QCD would give a vanishing contribution to the cosmological constant. 

\begin{itemize}
\item The Higgs boson could in principle be a dilaton e.g. \cite{Cata:2018wzl}
as it couples to mass via the compensator mechanism \eqref{eq:gSSD}. At leading
order in the low energy effective theory  it is equivalent to the coupling of the Higgs.
The basic idea is similar to technicolor (cf.
\cite{Hill:2002ap,Sannino:2009za} for reviews) in that a new gauge group is
added with techniquarks $q'$ which are in addition coupled to the weak force
such that the techniquark condensate breaks electroweak symmetry spontaneously,
this usually implies $F_\pi = v \approx 246 \GeV$. Whereas technicolor would be
classed as a Higgsless theory the same is not true in the dilaton case as it
takes on the role of the Higgs. Unlike technicolor the generation of fermion
mass terms is not aimed to be explained dynamically.

In our scenario the dilaton is a true GB in the $m_{q'} \to 0$ limit and
acquires its mass by explicit symmetry breaking 
\begin{equation}
  \label{eq:mH}
   m_D = \ORD(1) \, m_{q'}^\ym{1} \La'^{\ym{\ga_*}} \;,
\end{equation}
where $\La'$ is the hadronic scale of the new gauge sector.  \EQ\eqref{eq:mH}
suggests that a mass gap between $m_D$ and $\La'$ can be reached by making
$m_{q'}$ small. Whether or not $\La'$ can be sufficiently large,\footnote{By
large, we mean  larger than the naive estimate $\La' \approx 4 \pi F_\pi = 4 \pi
v \approx 3 \TeV$.} in order to avoid electroweak and LHC constraints, is
another question and beyond the scope of this paper. The crawling technicolor
scenario  in  \cite{Cata:2018wzl}, based on the dilaton, is different in that
the techniquarks are assumed to be massless and the dilaton/Higgs acquires its mass by
the hypothesis that  the IR fixed point is not (quite) reached. 
According to \cite{Cata:2018wzl} the dilaton mass
is then governed and made small by the derivative of the beta
function.

\item In this work, cf. \FIG\ref{fig:overview}, we have distinguished the
conformal dilaton phase from QCD but one might ask the question whether they are
one and the same.  Could it be that the so called $f_0(500)$ (cf.
\cite{Pelaez:2015qba} for a generic review on this particle), with pole on the
second sheet at $m_{f_0} = 449(20) - i \, 275(12)\, \MeV$~\cite{Zyla:2020zbs},
is a dilaton with mass $m_{f_0(500)} \propto \mq^{1/(1+\ga_*)}$?   
The first thing to note is that if one assumes the    Gell-Mann Oakes
Renner  relation, $F_\pi^2 m_{\pi}^2 = -  2 \mq \vev{\bar qq}$ (e.g. \cite{Donoghue:1992dd})
and $\vev{\bar qq} = \ORD(\La^3)$ then the mass scaling relation \eqref{eq:MassScaling2010}
 implies $\ga_* =1$. This is a logical possibility that is deserving of further studies.  
 In QCD, where $m_s \gg m_{u,d}$, 
 it is not immediate how to apply the mass scaling 
 relation \eqref{eq:MassScaling2010}.  The $f_0(500)$ surely has a strange quark component 
 and its mass scale can be considered to be of $\ORD(m_K)$. This is the case in terms of 
 the actual masses  
 and in scale chiral perturbation theory ~\cite{Crewther:2012wd,Crewther:2013vea,Crewther:2015dpa}.
 For more details about this EFT approach we refer the reader to 
  a series of works by Crewther and
Tunstall~\cite{Crewther:2012wd,Crewther:2013vea,Crewther:2015dpa}. Our approach is though different in that we consider the gluonic part proportional to the $\be$-function as subleading. 
Setting this aside, there are
interesting consequences for $K \to \pi\pi$ and the famous $\Delta I
=\frac{1}{2}$ rule. Such a scenario is also welcomed in dense nuclear interactions, combined 
with hidden local symmetry \cite{Rho:2021zwm,Brown:1991kk}.
\end{itemize}

\section{Discussions and Conclusions}
 \label{sec:con}
 
In this paper we have analysed the possibility of a conformal dilaton phase, in
addition to the QCD and conformal phase (cf. \TAB\ref{tab:overview} for comparison), where hadrons carry mass but the theory
is IR conformal. 
The mechanism  whereby this can happen is that conformal
symmetry is spontaneously broken and it is the corresponding Goldstone boson,
the dilaton, that restores the dilatation Ward identity~\eqref{eq:spin0con}.
More generally, we have shown, using renormalisation group arguments in
Euclidean space that the trace of the EMT vanishes on all physical states
\eqref{eq:imp}. This implies the vanishing of the gluon condensates and suggests 
that the scale breaking is driven by the quark condensate (cf. footnote \ref{foot:G2}).
This is an important result of our paper with consequences. For
example, it imposes an exact constraint on  the gravitational form
factors~\eqref{eq:spin0con}.
At zero momentum transfer we have shown that this constraint is satisfied,
\eqref{eq:new}, using the effective Lagrangian~\eqref{eq:gSSD}. As far as we
are aware this is a new result.

 \begin{table}[h]
	\centering
	\begin{tabular}{ l | c  c || l    l   l  }
   \text{phase} &  \text{GB}	  & $\matel{\phi}{\Tud{\mu}{\mu}}{\phi'}$  & $O_{had}$ 
   &  $(m_\dil^2,F_\dil)$  & $(m_\pi^2,F_\pi)$   \\ \hline
    \text{conformal window}    & $(-,-)$    &  $=0$   &    $ \mq^\frac{\eta}{1+\ga_*} $  & $(\mq^\frac{2}{1+\ga_*},\mq^\frac{1}{1+\ga_*}     )$  & $(\mq^\frac{2}{1+\ga_*},\mq^\frac{1}{1+\ga_*}     )$    \\
    \text{conformal dilaton}   &   $(\dil,\pi^a)$  &   $=0$ & $ \ORD(\Lambda) $ 
    & $(\mq^\frac{2}{1+\ga_*}, \ORD(\La)    )$  & $(\mq^\frac{2}{1+\ga_*} ,\ORD(\La)      )$    \\
    \text{QCD}  &  $(- ,\pi^a)$&   $\neq0$  &  $ \ORD(\Lambda) $
    & $( \ORD(\La^2) , \ORD(\La)    )$  & $(\mq ,\ORD(\La)      )$
	\end{tabular}
	\caption{\small Comparison table between the three different phases discussed in this paper. 
	GB stands for Goldstone boson. The states in the second column are physical states and zero quark 
	mass is assumed. The columns three to five indicate the scaling when a quark mass is turned on. 
	Above $O_{had}$ is a generic hadronic observable and 
	the conformal window scaling law has been discussed in  \eqref{eq:hyper}.
	 and the masses and the decay constant for the 
	Goldstone boson singled out in the least two columns.  
	The notation is the as in \TAB\ref{tab:summary} where further information on excited states in the 
	conformal dilaton phase can be found.}
	\label{tab:overview}
\end{table}

Such phases can be searched for in lattice Monte Carlo simulations for which we
have proposed  concrete signatures in \SEC\ref{sec:lattice}. First, the test of
the exact constraint on the gravitational form factor at zero momentum transfer
in \eqref{eq:new} and the scaling of the dilaton  mass \eqref{eq:mdisc}, as
compared to all other  ones. It will be interesting to see whether this new
perspective can resolve some of the debates  in the lattice conformal window
literature.

Moreover we have speculated in \SEC\ref{sec:dilaton} whether  a pseudo-dilaton
is present in QCD and or  the electroweak sector in terms of the $f_0(500)$ and the
Higgs boson. If the former were true then this would suggest that the conformal
dilaton- and the QCD-phase are one and the same.\footnote{IR conformality is
important for renormalisation group flow theorems e.g. in the topological Euler
term~\cite{Komargodski:2011vj,Luty:2012ww}, and the $\Box R$
term~\cite{Prochazka:2017pfa}.} In our view the study of whether the Higgs is a
composite dilaton is deserving of further attention 
also because it has the potential to ameliorate the cosmological 
constant problem as emphasised earlier.

Finally we wish to comment on the different (pseudo) Goldstone bosons,
the pions, the $\eta'$ and the dilaton associated with breaking of the
$SU_A(N_f)$, the $U_A(1)$ and the dilatation symmetry. In all three cases the
quark masses are a form of explicit symmetry breaking. This is manifested by the
corresponding WIs \eqref{eq:DivNoetherCurrTwo} and  
$\partial \cdot J_5 = 2 \mq P + \frac{g^2}{16 \pi^2} G \tilde{G}$ for the
$U_A(1)$-case. The axial non-singlet case stands  out in that there is no
anomalous piece and it is indeed the case, as well-known, that in the $\mq \to
0$ limit the pions become true Goldstone bosons.  
In the axial singlet case the anomalous piece does contribute to the large
$\eta'$ mass which constitutes the resolution to the $U_A(1)$-problem
\cite{tHooft:1976rip}). It is noted that the $\eta'$ becomes a Goldstone boson
in the  $N_c \to \infty$ limit as the anomalous term is $1/N_c$ suppressed.  On
the other hand, according to our analysis, the anomalous breaking of the scale
symmetry does not affect the dilaton mass in the $m_q \to 0$ limit and it thus
is, remarkably,  a genuine Goldstone boson.

\acknowledgments
LDD and RZ are supported by an STFC Consolidated Grant, ST/P0000630/1. We are
grateful to Tyler Corbett,  Daniel Litim,  Oliver Witzel and especially to  Lewis Tunstall for
very useful discussions on the topic. Furthermore we thank Marc Knecht for pointing out a minor 
typo in \eqref{eq:spin0con} with subsequent amendment in \eqref{eq:crucial}.

\appendix

\section{Conventions}
\label{app:conv}

We write the gauge theory Lagrangian as
\begin{equation}
\label{eq:QCD}
{\cal L} = - \frac{1}{4}G^2  + N_f \mq \bar q (i \slashed{D} -\mq) q \;,
\end{equation}
where $G^2 = G_{\mu\nu}^A G^{A \mu\nu}$ is the field strength tensor squared and
$A$ is the adjoint index of the gauge group, $N_f$ is the number of flavours,
which we assume to have degenerate mass $m_q$. The beta function and anomalous
dimension are defined by ${\be} = \frac{d}{d\ln \mu}   g $ and $\ga  = -
\frac{d}{d\ln \mu} \ln \mq $. Our Minkowski metric is $\mink_{\mu \nu} =
\textrm{diag}(1,-1,-1,-1)$.

\section{Gravitational Form Factors of Spin-$1/2$} 
\label{app:12}

In this appendix we present the spin-$1/2$ case, completing the spin-$0$ form
factor discussion in the main text. The analogous definition of the scalar case
\eqref{eq:GF}, denoting the spin-$1/2$ fermion by $\bary$, is given by
\begin{alignat}{2}
  \label{eq:GF12}
  & \Tbary_{\mu \nu}(p,p') &\;=\;& 
    \matel{\bary(p')}{T_{\mu\nu}(0)}{\bary(p)}    \\[0.1cm]
  &  &\;=\;&  \bar u(p') \left(  \frac{1}{2} 
    \ga_{\{ \mu} {\cal P}_{\nu\}} \, g_1(q^2) +  
    \frac{i  {\cal P}_{\{ \mu } \sig_{\nu\}q }}{4 m_{\bary}} \,  {g_2(q^2)} +   
    \left(\frac {q_\mu q_\nu}{q^2}-  \mink_{\mu \nu} \right) 
  \frac{m_{\bary }}{4}  \,g_3(q^2)  \right) u(p) \;,
    \nonumber
\end{alignat}
where $\sig_{\mu q} = \sig_{\mu\nu}q^\nu$. Now, $g_1(0) = 1$ in order to get the
mass relation correct and $g_2(0) = 0$ is equivalent to the vanishing of the
nucleon's gravitomagnetic moment \cite{Hudson:2016gnq} ($(g_1,g_2,g_3) =
(A,B,D \frac{q^2}{m_{\bary}^2})$ in their notation). The third form factor, sometimes referred to as the
$D$-term but denoted by $g_3$ in order to avoid confusion with the dilaton, is
unknown although related to the pressure and the shear.  

The above remarks are general and we now turn to the conformal dilaton scenario.
It is the  $g_3$ term that plays the analogue role with respect to the dilaton
pole. For $ \Tbaryud{\mu}{\mu}(p,p)=0$, $m_{\bary}^2 \neq 0$  to be true
the following constraint on the form factors has to hold
\begin{equation}
  \label{eq:spin12con}
  m_{\bary}  g_1(q^2) - \frac{q^2}{8 m_{\bary}} g_2(q^2)  - 
  \frac{(d-1)}{4 } m_{\bary} g_3(q^2) =0 \;.
\end{equation}
Since $g_1(0)=1$ and $g_2(0) = 0$ we must have that
\begin{equation}
  \label{eq:require}
   g_3(0) = 
  \frac{4 }{d-1} \;,
\end{equation}
 holds in analogy with \eqref{eq:crucial} or more precisely
\eqref{eq:spin0con}.  It then remains to check that this is indeed the case from
the hadronic viewpoint. It is readily verified that $g_{\bary\bary \dil} =
\frac{m_{\bary}}{F_{\dil}}$ using the same steps as in \eqref{eq:gSSD} with 
\begin{equation}
  \label{eq:Dff}
  \vev{\dil \bary | \bary } =  i (2\pi)^d \de\left(\sum p_i\right)\, 
  g_{\bary \bary \dil } \bar u(p') u(p)\;.
 \end{equation}
In analogy to \eqref{eq:analogy1} we have
\begin{alignat}{2}
  \label{eq:Dff2}
  & \vev{\dil \bary | \bary } &\;=\;& 
    \lim_{q^2 \to 0} i \frac{q^2}{Z'_{\dil}} 
    \int d^d x e^{i q\cdot x}  \Tbary_{33}(p,p',x)   \nonumber \\[0.1cm]
  & &\;=\;&  \lim_{q^2 \to 0} i \frac{q^2}{Z'_{\dil}}   \bar u(p')\frac{m_f^2}{q^2} g_3(q^2) u(p)(2\pi)^d  \de(\sum p_i)  \;,
\end{alignat}
where here $Z_D' = 4 F_D m_{\bary}/(d-1)$ is the appropriate LSZ factor.
Equating \eqref{eq:Dff} and \eqref{eq:Dff2} one finally gets
\begin{equation}
 g_3(0) =  
 \frac{g_{\bary \bary \dil }   Z_D' }{m_{\bary}^2} =  \frac{4  }{d-1} \;,
\end{equation}
as required by \eqref{eq:require}.

\bibliographystyle{utphys}
\bibliography{Dil-refs.bib}

\providecommand{\href}[2]{#2}\begingroup\raggedright\begin{thebibliography}{10}

\bibitem{Banks:1981nn}
T.~Banks and A.~Zaks, ``{On the Phase Structure of Vector-Like Gauge Theories
  with Massless Fermions},''
  \href{http://dx.doi.org/10.1016/0550-3213(82)90035-9}{{\em Nucl. Phys. B}
  {\bfseries 196} (1982) 189--204}.

\bibitem{Seiberg:1994bz}
N.~Seiberg, ``{Exact results on the space of vacua of four-dimensional SUSY
  gauge theories},'' \href{http://dx.doi.org/10.1103/PhysRevD.49.6857}{{\em
  Phys. Rev. D} {\bfseries 49} (1994) 6857--6863},
  \href{http://arxiv.org/abs/hep-th/9402044}{{\ttfamily arXiv:hep-th/9402044}}.

\bibitem{Terning:2006bq}
J.~Terning,
  \href{http://dx.doi.org/10.1093/acprof:oso/9780198567639.001.0001}{{\em
  {Modern supersymmetry: Dynamics and duality}}}.
\newblock 2006.

\bibitem{Nogradi:2016qek}
D.~Nogradi and A.~Patella, ``{Strong dynamics, composite Higgs and the
  conformal window},'' \href{http://dx.doi.org/10.1142/S0217751X1643003X}{{\em
  Int. J. Mod. Phys. A} {\bfseries 31} no.~22, (2016) 1643003},
  \href{http://arxiv.org/abs/1607.07638}{{\ttfamily arXiv:1607.07638
  [hep-lat]}}.

\bibitem{Nakayama:2013is}
Y.~Nakayama, ``{Scale invariance vs conformal invariance},''
  \href{http://dx.doi.org/10.1016/j.physrep.2014.12.003}{{\em Phys. Rept.}
  {\bfseries 569} (2015) 1--93},
  \href{http://arxiv.org/abs/1302.0884}{{\ttfamily arXiv:1302.0884 [hep-th]}}.

\bibitem{Yamawaki:1985zg}
K.~Yamawaki, M.~Bando, and K.-i. Matumoto, ``{Scale Invariant Technicolor Model
  and a Technidilaton},''
  \href{http://dx.doi.org/10.1103/PhysRevLett.56.1335}{{\em Phys. Rev. Lett.}
  {\bfseries 56} (1986) 1335}.

\bibitem{Cata:2018wzl}
O.~Cat\`a, R.~J. Crewther, and L.~C. Tunstall, ``{Crawling technicolor},''
  \href{http://dx.doi.org/10.1103/PhysRevD.100.095007}{{\em Phys. Rev. D}
  {\bfseries 100} no.~9, (2019) 095007},
  \href{http://arxiv.org/abs/1803.08513}{{\ttfamily arXiv:1803.08513
  [hep-ph]}}.

\bibitem{Appelquist:2020bqj}
T.~Appelquist, J.~Ingoldby, and M.~Piai, ``{Nearly Conformal Composite Higgs
  Model},'' \href{http://dx.doi.org/10.1103/PhysRevLett.126.191804}{{\em Phys.
  Rev. Lett.} {\bfseries 126} no.~19, (2021) 191804},
  \href{http://arxiv.org/abs/2012.09698}{{\ttfamily arXiv:2012.09698
  [hep-ph]}}.

\bibitem{Appelquist:2017wcg}
T.~Appelquist, J.~Ingoldby, and M.~Piai, ``{Dilaton EFT Framework For Lattice
  Data},'' \href{http://dx.doi.org/10.1007/JHEP07(2017)035}{{\em JHEP}
  {\bfseries 07} (2017) 035}, \href{http://arxiv.org/abs/1702.04410}{{\ttfamily
  arXiv:1702.04410 [hep-ph]}}.

\bibitem{Appelquist:2017vyy}
T.~Appelquist, J.~Ingoldby, and M.~Piai, ``{Analysis of a Dilaton EFT for
  Lattice Data},'' \href{http://dx.doi.org/10.1007/JHEP03(2018)039}{{\em JHEP}
  {\bfseries 03} (2018) 039}, \href{http://arxiv.org/abs/1711.00067}{{\ttfamily
  arXiv:1711.00067 [hep-ph]}}.

\bibitem{Golterman:2016lsd}
M.~Golterman and Y.~Shamir, ``{Low-energy effective action for pions and a
  dilatonic meson},'' \href{http://dx.doi.org/10.1103/PhysRevD.94.054502}{{\em
  Phys. Rev. D} {\bfseries 94} no.~5, (2016) 054502},
  \href{http://arxiv.org/abs/1603.04575}{{\ttfamily arXiv:1603.04575
  [hep-ph]}}.

\bibitem{Matsuzaki:2013eva}
S.~Matsuzaki and K.~Yamawaki, ``{Dilaton Chiral Perturbation Theory:
  Determining the Mass and Decay Constant of the Technidilaton on the
  Lattice},'' \href{http://dx.doi.org/10.1103/PhysRevLett.113.082002}{{\em
  Phys. Rev. Lett.} {\bfseries 113} no.~8, (2014) 082002},
  \href{http://arxiv.org/abs/1311.3784}{{\ttfamily arXiv:1311.3784 [hep-lat]}}.

\bibitem{Csaki:2014bua}
C.~Csaki, N.~Kaloper, J.~Serra, and J.~Terning, ``{Inflation from Broken Scale
  Invariance},'' \href{http://dx.doi.org/10.1103/PhysRevLett.113.161302}{{\em
  Phys. Rev. Lett.} {\bfseries 113} (2014) 161302},
  \href{http://arxiv.org/abs/1406.5192}{{\ttfamily arXiv:1406.5192 [hep-th]}}.

\bibitem{Minkowski:1976en}
P.~Minkowski, ``{On the Anomalous Divergence of the Dilatation Current in Gauge
  Theories},''.

\bibitem{Collins:1976yq}
J.~C. Collins, A.~Duncan, and S.~D. Joglekar, ``{Trace and Dilatation Anomalies
  in Gauge Theories},''
\href{http://dx.doi.org/10.1103/PhysRevD.16.438}{{\em Phys. Rev.} {\bfseries
  D16} (1977) 438--449}.

\bibitem{Nielsen:1977sy}
N.~K. Nielsen, ``{The Energy Momentum Tensor in a Nonabelian Quark Gluon
  Theory},''
\href{http://dx.doi.org/10.1016/0550-3213(77)90040-2}{{\em Nucl. Phys.}
  {\bfseries B120} (1977) 212--220}.

\bibitem{Goldstone:1962es}
J.~Goldstone, A.~Salam, and S.~Weinberg, ``{Broken Symmetries},''
  \href{http://dx.doi.org/10.1103/PhysRev.127.965}{{\em Phys. Rev.} {\bfseries
  127} (1962) 965--970}.

\bibitem{Baikov:2016tgj}
P.~A. Baikov, K.~G. Chetyrkin, and J.~H. K\"uhn, ``{Five-Loop Running of the
  QCD coupling constant},''
  \href{http://dx.doi.org/10.1103/PhysRevLett.118.082002}{{\em Phys. Rev.
  Lett.} {\bfseries 118} no.~8, (2017) 082002},
  \href{http://arxiv.org/abs/1606.08659}{{\ttfamily arXiv:1606.08659
  [hep-ph]}}.

\bibitem{Ryttov:2010iz}
T.~A. Ryttov and R.~Shrock, ``{Higher-Loop Corrections to the Infrared
  Evolution of a Gauge Theory with Fermions},''
  \href{http://dx.doi.org/10.1103/PhysRevD.83.056011}{{\em Phys. Rev. D}
  {\bfseries 83} (2011) 056011},
  \href{http://arxiv.org/abs/1011.4542}{{\ttfamily arXiv:1011.4542 [hep-ph]}}.

\bibitem{Ryttov:2016ner}
T.~A. Ryttov and R.~Shrock, ``{Infrared Zero of $\beta$ and Value of $\gamma_m$
  for an SU(3) Gauge Theory at the Five-Loop Level},''
  \href{http://dx.doi.org/10.1103/PhysRevD.94.105015}{{\em Phys. Rev. D}
  {\bfseries 94} no.~10, (2016) 105015},
  \href{http://arxiv.org/abs/1607.06866}{{\ttfamily arXiv:1607.06866
  [hep-th]}}.

\bibitem{Ryttov:2016hal}
T.~A. Ryttov and R.~Shrock, ``{Scheme-Independent Series Expansions at an
  Infrared Zero of the Beta Function in Asymptotically Free Gauge Theories},''
  \href{http://dx.doi.org/10.1103/PhysRevD.94.125005}{{\em Phys. Rev. D}
  {\bfseries 94} no.~12, (2016) 125005},
  \href{http://arxiv.org/abs/1610.00387}{{\ttfamily arXiv:1610.00387
  [hep-th]}}.

\bibitem{Hasenfratz:2020ess}
A.~Hasenfratz, C.~Rebbi, and O.~Witzel, ``{Gradient flow step-scaling function
  for SU(3) with ten fundamental flavors},''
  \href{http://dx.doi.org/10.1103/PhysRevD.101.114508}{{\em Phys. Rev. D}
  {\bfseries 101} no.~11, (2020) 114508},
  \href{http://arxiv.org/abs/2004.00754}{{\ttfamily arXiv:2004.00754
  [hep-lat]}}.

\bibitem{Fodor:2019ypi}
Z.~Fodor, K.~Holland, J.~Kuti, D.~Nogradi, and C.~H. Wong, ``{Case studies of
  near-conformal $\beta$-functions},''
  \href{http://dx.doi.org/10.22323/1.363.0121}{{\em PoS} {\bfseries
  LATTICE2019} (2019) 121}, \href{http://arxiv.org/abs/1912.07653}{{\ttfamily
  arXiv:1912.07653 [hep-lat]}}.

\bibitem{Fodor:2017gtj}
Z.~Fodor, K.~Holland, J.~Kuti, D.~Nogradi, and C.~H. Wong, ``{Extended
  investigation of the twelve-flavor $\beta$-function},''
  \href{http://dx.doi.org/10.1016/j.physletb.2018.02.008}{{\em Phys. Lett. B}
  {\bfseries 779} (2018) 230--236},
  \href{http://arxiv.org/abs/1710.09262}{{\ttfamily arXiv:1710.09262
  [hep-lat]}}.

\bibitem{Fodor:2018uih}
Z.~Fodor, K.~Holland, J.~Kuti, D.~Nogradi, and C.~H. Wong, ``{Is SU(3) gauge
  theory with 13 massless flavors conformal?},''
  \href{http://dx.doi.org/10.22323/1.334.0198}{{\em PoS} {\bfseries
  LATTICE2018} (2018) 198}, \href{http://arxiv.org/abs/1811.05024}{{\ttfamily
  arXiv:1811.05024 [hep-lat]}}.

\bibitem{Chiu:2018edw}
T.-W. Chiu, ``{Improved study of the $\beta$-function of $SU(3)$ gauge theory
  with $N_f = 10$ massless domain-wall fermions},''
  \href{http://dx.doi.org/10.1103/PhysRevD.99.014507}{{\em Phys. Rev. D}
  {\bfseries 99} no.~1, (2019) 014507},
  \href{http://arxiv.org/abs/1811.01729}{{\ttfamily arXiv:1811.01729
  [hep-lat]}}.

\bibitem{LatticeStrongDynamics:2018hun}
{\bfseries Lattice Strong Dynamics} Collaboration, T.~Appelquist {\em et~al.},
  ``{Nonperturbative investigations of SU(3) gauge theory with eight dynamical
  flavors},'' \href{http://dx.doi.org/10.1103/PhysRevD.99.014509}{{\em Phys.
  Rev. D} {\bfseries 99} no.~1, (2019) 014509},
  \href{http://arxiv.org/abs/1807.08411}{{\ttfamily arXiv:1807.08411
  [hep-lat]}}.

\bibitem{DelDebbio:2015byq}
L.~Del~Debbio, B.~Lucini, A.~Patella, C.~Pica, and A.~Rago, ``{Large volumes
  and spectroscopy of walking theories},''
  \href{http://dx.doi.org/10.1103/PhysRevD.93.054505}{{\em Phys. Rev. D}
  {\bfseries 93} no.~5, (2016) 054505},
  \href{http://arxiv.org/abs/1512.08242}{{\ttfamily arXiv:1512.08242
  [hep-lat]}}.

\bibitem{Brower:2015owo}
R.~C. Brower, A.~Hasenfratz, C.~Rebbi, E.~Weinberg, and O.~Witzel, ``{Composite
  Higgs model at a conformal fixed point},''
  \href{http://dx.doi.org/10.1103/PhysRevD.93.075028}{{\em Phys. Rev. D}
  {\bfseries 93} no.~7, (2016) 075028},
  \href{http://arxiv.org/abs/1512.02576}{{\ttfamily arXiv:1512.02576
  [hep-ph]}}.

\bibitem{LSD:2014nmn}
{\bfseries LSD} Collaboration, T.~Appelquist {\em et~al.}, ``{Lattice
  simulations with eight flavors of domain wall fermions in SU(3) gauge
  theory},'' \href{http://dx.doi.org/10.1103/PhysRevD.90.114502}{{\em Phys.
  Rev. D} {\bfseries 90} no.~11, (2014) 114502},
  \href{http://arxiv.org/abs/1405.4752}{{\ttfamily arXiv:1405.4752 [hep-lat]}}.

\bibitem{LatKMI:2014xoh}
{\bfseries LatKMI} Collaboration, Y.~Aoki {\em et~al.}, ``{Light composite
  scalar in eight-flavor QCD on the lattice},''
  \href{http://dx.doi.org/10.1103/PhysRevD.89.111502}{{\em Phys. Rev. D}
  {\bfseries 89} (2014) 111502},
  \href{http://arxiv.org/abs/1403.5000}{{\ttfamily arXiv:1403.5000 [hep-lat]}}.

\bibitem{LatKMI:2016xxi}
{\bfseries LatKMI} Collaboration, Y.~Aoki {\em et~al.}, ``{Light flavor-singlet
  scalars and walking signals in $N_f=8$ QCD on the lattice},''
  \href{http://dx.doi.org/10.1103/PhysRevD.96.014508}{{\em Phys. Rev. D}
  {\bfseries 96} no.~1, (2017) 014508},
  \href{http://arxiv.org/abs/1610.07011}{{\ttfamily arXiv:1610.07011
  [hep-lat]}}.

\bibitem{Witzel:2019jbe}
O.~Witzel, ``{Review on Composite Higgs Models},''
  \href{http://dx.doi.org/10.22323/1.334.0006}{{\em PoS} {\bfseries
  LATTICE2018} (2019) 006}, \href{http://arxiv.org/abs/1901.08216}{{\ttfamily
  arXiv:1901.08216 [hep-lat]}}.

\bibitem{Holstein:2006ud}
B.~R. Holstein, ``{Metric modifications for a massive spin 1 particle},''
  \href{http://dx.doi.org/10.1103/PhysRevD.74.084030}{{\em Phys. Rev. D}
  {\bfseries 74} (2006) 084030},
  \href{http://arxiv.org/abs/gr-qc/0607051}{{\ttfamily arXiv:gr-qc/0607051}}.

\bibitem{Leader:2013jra}
E.~Leader and C.~Lorc\'e, ``{The angular momentum controversy:
  What\textquoteright{}s it all about and does it matter?},''
  \href{http://dx.doi.org/10.1016/j.physrep.2014.02.010}{{\em Phys. Rept.}
  {\bfseries 541} no.~3, (2014) 163--248},
  \href{http://arxiv.org/abs/1309.4235}{{\ttfamily arXiv:1309.4235 [hep-ph]}}.

\bibitem{Hudson:2016gnq}
J.~Hudson, I.~A. Perevalova, M.~V. Polyakov, and P.~Schweitzer, ``{Structure of
  the Energy-Momentum Tensor and Applications},''
  \href{http://dx.doi.org/10.22323/1.284.0007}{{\em PoS} {\bfseries QCDEV2016}
  (2017) 007}, \href{http://arxiv.org/abs/1612.06721}{{\ttfamily
  arXiv:1612.06721 [hep-ph]}}.

\bibitem{Callan:1970ze}
C.~G. Callan, Jr., S.~R. Coleman, and R.~Jackiw, ``{A New improved energy -
  momentum tensor},''
  \href{http://dx.doi.org/10.1016/0003-4916(70)90394-5}{{\em Annals Phys.}
  {\bfseries 59} (1970) 42--73}.

\bibitem{Donoghue:1992dd}
J.~F. Donoghue, E.~Golowich, and B.~R. Holstein,
  \href{http://dx.doi.org/10.1017/CBO9780511524370}{{\em {Dynamics of the
  standard model}}}, vol.~2.
\newblock CUP, 2014.

\bibitem{Weinberg:1980kq}
S.~Weinberg and E.~Witten, ``{Limits on Massless Particles},''
  \href{http://dx.doi.org/10.1016/0370-2693(80)90212-9}{{\em Phys. Lett. B}
  {\bfseries 96} (1980) 59--62}.

\bibitem{DelDebbio:2013sta}
L.~Del~Debbio and R.~Zwicky, ``{Renormalisation group, trace anomaly and
  Feynman\textendash{}Hellmann theorem},''
  \href{http://dx.doi.org/10.1016/j.physletb.2014.05.038}{{\em Phys. Lett. B}
  {\bfseries 734} (2014) 107--110},
  \href{http://arxiv.org/abs/1306.4274}{{\ttfamily arXiv:1306.4274 [hep-ph]}}.

\bibitem{Gell-Mann:1969rik}
M.~Gell-Mann, \href{http://dx.doi.org/10.2172/4147223}{{\em {Symmetry violation
  in hadron physics}}}.
\newblock 1969.

\bibitem{Weinberg:1995mt}
S.~Weinberg, {\em {The Quantum theory of fields. Vol. 1: Foundations}}.
\newblock Cambridge University Press, 6, 2005.

\bibitem{Duncan:2012aja}
A.~Duncan,
  \href{http://dx.doi.org/10.1093/acprof:oso/9780199573264.001.0001}{{\em {The
  Conceptual Framework of Quantum Field Theory}}}.
\newblock Oxford University Press, 8, 2012.

\bibitem{Coleman:1985rnk}
S.~Coleman, \href{http://dx.doi.org/10.1017/CBO9780511565045}{{\em {Aspects of
  Symmetry}: {Selected Erice Lectures}}}.
\newblock Cambridge University Press, Cambridge, U.K., 1985.

\bibitem{DelDebbio:2010ze}
L.~Del~Debbio and R.~Zwicky, ``{Hyperscaling relations in mass-deformed
  conformal gauge theories},''
  \href{http://dx.doi.org/10.1103/PhysRevD.82.014502}{{\em Phys. Rev. D}
  {\bfseries 82} (2010) 014502},
  \href{http://arxiv.org/abs/1005.2371}{{\ttfamily arXiv:1005.2371 [hep-ph]}}.

\bibitem{DelDebbio:2010jy}
L.~Del~Debbio and R.~Zwicky, ``{Scaling relations for the entire spectrum in
  mass-deformed conformal gauge theories},''
  \href{http://dx.doi.org/10.1016/j.physletb.2011.04.059}{{\em Phys. Lett. B}
  {\bfseries 700} (2011) 217--220},
  \href{http://arxiv.org/abs/1009.2894}{{\ttfamily arXiv:1009.2894 [hep-ph]}}.

\bibitem{DelDebbio:2013qta}
L.~Del~Debbio and R.~Zwicky, ``{Conformal scaling and the size of
  $m$-hadrons},'' \href{http://dx.doi.org/10.1103/PhysRevD.89.014503}{{\em
  Phys. Rev. D} {\bfseries 89} no.~1, (2014) 014503},
  \href{http://arxiv.org/abs/1306.4038}{{\ttfamily arXiv:1306.4038 [hep-ph]}}.

\bibitem{Komargodski:2011vj}
Z.~Komargodski and A.~Schwimmer, ``{On Renormalization Group Flows in Four
  Dimensions},'' \href{http://dx.doi.org/10.1007/JHEP12(2011)099}{{\em JHEP}
  {\bfseries 12} (2011) 099}, \href{http://arxiv.org/abs/1107.3987}{{\ttfamily
  arXiv:1107.3987 [hep-th]}}.

\bibitem{Shore:2016xor}
G.~M. Shore, \href{http://dx.doi.org/10.1007/978-3-319-54000-9}{{\em {The c and
  a-theorems and the Local Renormalisation Group}}}.
\newblock SpringerBriefs in Physics. Springer, Cham, 2017.
\newblock \href{http://arxiv.org/abs/1601.06662}{{\ttfamily arXiv:1601.06662
  [hep-th]}}.

\bibitem{Prochazka:2017pfa}
V.~Prochazka and R.~Zwicky, ``{On the Flow of $\Box R$ Weyl-Anomaly},''
  \href{http://dx.doi.org/10.1103/PhysRevD.96.045011}{{\em Phys. Rev. D}
  {\bfseries 96} no.~4, (2017) 045011},
  \href{http://arxiv.org/abs/1703.01239}{{\ttfamily arXiv:1703.01239
  [hep-th]}}.

\bibitem{Weinberg:1996kr}
S.~Weinberg, {\em {The quantum theory of fields. Vol. 2: Modern applications}}.
\newblock Cambridge University Press, 8, 2013.

\bibitem{Hill:2002ap}
C.~T. Hill and E.~H. Simmons, ``{Strong Dynamics and Electroweak Symmetry
  Breaking},'' \href{http://dx.doi.org/10.1016/S0370-1573(03)00140-6}{{\em
  Phys. Rept.} {\bfseries 381} (2003) 235--402},
  \href{http://arxiv.org/abs/hep-ph/0203079}{{\ttfamily arXiv:hep-ph/0203079}}.
  [Erratum: Phys.Rept. 390, 553--554 (2004)].

\bibitem{Sannino:2009za}
F.~Sannino, ``{Conformal Dynamics for TeV Physics and Cosmology},'' {\em Acta
  Phys. Polon. B} {\bfseries 40} (2009) 3533--3743,
  \href{http://arxiv.org/abs/0911.0931}{{\ttfamily arXiv:0911.0931 [hep-ph]}}.

\bibitem{Pelaez:2015qba}
J.~R. Pelaez, ``{From controversy to precision on the sigma meson: a review on
  the status of the non-ordinary $f_0(500)$ resonance},''
  \href{http://dx.doi.org/10.1016/j.physrep.2016.09.001}{{\em Phys. Rept.}
  {\bfseries 658} (2016) 1}, \href{http://arxiv.org/abs/1510.00653}{{\ttfamily
  arXiv:1510.00653 [hep-ph]}}.

\bibitem{Zyla:2020zbs}
{\bfseries Particle Data Group} Collaboration, P.~Zyla {\em et~al.}, ``{Review
  of Particle Physics},'' \href{http://dx.doi.org/10.1093/ptep/ptaa104}{{\em
  PTEP} {\bfseries 2020} no.~8, (2020) 083C01}.

\bibitem{Crewther:2012wd}
R.~J. Crewther and L.~C. Tunstall, ``{Origin of $\Delta I=1/2$ Rule for Kaon
  Decays: QCD Infrared Fixed Point},''
  \href{http://arxiv.org/abs/1203.1321}{{\ttfamily arXiv:1203.1321 [hep-ph]}}.

\bibitem{Crewther:2013vea}
R.~J. Crewther and L.~C. Tunstall, ``{$\Delta I=1/2$ rule for kaon decays
  derived from QCD infrared fixed point},''
  \href{http://dx.doi.org/10.1103/PhysRevD.91.034016}{{\em Phys. Rev. D}
  {\bfseries 91} no.~3, (2015) 034016},
  \href{http://arxiv.org/abs/1312.3319}{{\ttfamily arXiv:1312.3319 [hep-ph]}}.

\bibitem{Crewther:2015dpa}
R.~J. Crewther and L.~C. Tunstall, ``{Status of Chiral-Scale Perturbation
  Theory},'' \href{http://dx.doi.org/10.22323/1.253.0132}{{\em PoS} {\bfseries
  CD15} (2015) 132}, \href{http://arxiv.org/abs/1510.01322}{{\ttfamily
  arXiv:1510.01322 [hep-ph]}}.

\bibitem{Rho:2021zwm}
M.~Rho and Y.-L. Ma, ``{Manifestation of Hidden Symmetries in Baryonic Matter:
  From Finite Nuclei to Neutron Stars},''
  \href{http://dx.doi.org/10.1142/S0217732321300123}{{\em Mod. Phys. Lett. A}
  {\bfseries 36} no.~13, (2021) 2130012},
  \href{http://arxiv.org/abs/2101.07121}{{\ttfamily arXiv:2101.07121
  [nucl-th]}}.

\bibitem{Brown:1991kk}
G.~E. Brown and M.~Rho, ``{Scaling effective Lagrangians in a dense medium},''
  \href{http://dx.doi.org/10.1103/PhysRevLett.66.2720}{{\em Phys. Rev. Lett.}
  {\bfseries 66} (1991) 2720--2723}.

\bibitem{Luty:2012ww}
M.~A. Luty, J.~Polchinski, and R.~Rattazzi, ``{The $a$-theorem and the
  Asymptotics of 4D Quantum Field Theory},''
  \href{http://dx.doi.org/10.1007/JHEP01(2013)152}{{\em JHEP} {\bfseries 01}
  (2013) 152}, \href{http://arxiv.org/abs/1204.5221}{{\ttfamily arXiv:1204.5221
  [hep-th]}}.

\bibitem{tHooft:1976rip}
G.~'t~Hooft, ``{Symmetry Breaking Through Bell-Jackiw Anomalies},''
  \href{http://dx.doi.org/10.1103/PhysRevLett.37.8}{{\em Phys. Rev. Lett.}
  {\bfseries 37} (1976) 8--11}.

\end{thebibliography}\endgroup

\end{document}